\PassOptionsToPackage{unicode}{hyperref}
\PassOptionsToPackage{hyphens}{url}
\PassOptionsToPackage{dvipsnames,svgnames,x11names}{xcolor}
\documentclass[
  12pt]{article}

\usepackage{amsmath,amssymb}
\usepackage{iftex}
\usepackage{threeparttable}
\usepackage{amsmath}
\ifPDFTeX
\usepackage{enumitem}
  \usepackage[T1]{fontenc}
  \usepackage[utf8]{inputenc}
  \usepackage{textcomp} 
\else 
  \usepackage{unicode-math}
  \defaultfontfeatures{Scale=MatchLowercase}
  \defaultfontfeatures[\rmfamily]{Ligatures=TeX,Scale=1}
\fi
\usepackage{lmodern}
\ifPDFTeX\else  
\fi
\IfFileExists{upquote.sty}{\usepackage{upquote}}{}
\IfFileExists{microtype.sty}{
  \usepackage[]{microtype}
  \UseMicrotypeSet[protrusion]{basicmath} 
}{}
\makeatletter
\@ifundefined{KOMAClassName}{
  \IfFileExists{parskip.sty}{%
    \usepackage{parskip}
  }{
    \setlength{\parindent}{0pt}
    \setlength{\parskip}{6pt plus 2pt minus 1pt}}
}{
  \KOMAoptions{parskip=half}}
\makeatother
\usepackage{xcolor}
\makeatletter
\ifx\paragraph\undefined\else
  \let\oldparagraph\paragraph
  \renewcommand{\paragraph}{
    \@ifstar
      \xxxParagraphStar
      \xxxParagraphNoStar
  }
  \newcommand{\xxxParagraphStar}[1]{\oldparagraph*{#1}\mbox{}}
  \newcommand{\xxxParagraphNoStar}[1]{\oldparagraph{#1}\mbox{}}
\fi
\ifx\subparagraph\undefined\else
  \let\oldsubparagraph\subparagraph
  \renewcommand{\subparagraph}{
    \@ifstar
      \xxxSubParagraphStar
      \xxxSubParagraphNoStar
  }
  \newcommand{\xxxSubParagraphStar}[1]{\oldsubparagraph*{#1}\mbox{}}
  \newcommand{\xxxSubParagraphNoStar}[1]{\oldsubparagraph{#1}\mbox{}}
\fi
\makeatother

\usepackage{longtable,booktabs,array}
\usepackage{calc} 
\usepackage{etoolbox}
\makeatletter
\patchcmd\longtable{\par}{\if@noskipsec\mbox{}\fi\par}{}{}
\makeatother
\IfFileExists{footnotehyper.sty}{\usepackage{footnotehyper}}{\usepackage{footnote}}
\makesavenoteenv{longtable}
\usepackage{graphicx}
\makeatletter
\def\maxwidth{\ifdim\Gin@nat@width>\linewidth\linewidth\else\Gin@nat@width\fi}
\def\maxheight{\ifdim\Gin@nat@height>\textheight\textheight\else\Gin@nat@height\fi}
\makeatother
\setkeys{Gin}{width=\maxwidth,height=\maxheight,keepaspectratio}
\makeatletter
\def\fps@figure{htbp}
\makeatother

\makeatletter
\@ifpackageloaded{caption}{}{\usepackage{caption}}
\AtBeginDocument{%
\ifdefined\contentsname
  \renewcommand*\contentsname{Table of contents}
\else
  \newcommand\contentsname{Table of contents}
\fi
\ifdefined\listfigurename
  \renewcommand*\listfigurename{List of Figures}
\else
  \newcommand\listfigurename{List of Figures}
\fi
\ifdefined\listtablename
  \renewcommand*\listtablename{List of Tables}
\else
  \newcommand\listtablename{List of Tables}
\fi
\ifdefined\figurename
  \renewcommand*\figurename{Figure}
\else
  \newcommand\figurename{Figure}
\fi
\ifdefined\tablename
  \renewcommand*\tablename{Table}
\else
  \newcommand\tablename{Table}
\fi
}
\@ifpackageloaded{float}{}{\usepackage{float}}
\floatstyle{ruled}
\@ifundefined{c@chapter}{\newfloat{codelisting}{h}{lop}}{\newfloat{codelisting}{h}{lop}[chapter]}
\floatname{codelisting}{Listing}

\makeatother
\makeatletter
\@ifpackageloaded{caption}{}{\usepackage{caption}}
\@ifpackageloaded{subcaption}{}{\usepackage{subcaption}}
\makeatother

\ifLuaTeX
  \usepackage{selnolig}  
\fi
\usepackage[]{natbib}

\usepackage{bookmark}

\IfFileExists{xurl.sty}{\usepackage{xurl}}{} 
\hypersetup{
  pdftitle={Title},
  pdfauthor={Author 1; Author 2},
  pdfkeywords={3 to 6 keywords, that do not appear in the title},
  colorlinks=true,
  linkcolor={blue},
  filecolor={Maroon},
  citecolor={Blue},
  urlcolor={Blue},
  pdfcreator={LaTeX via pandoc}}

\newtheorem{assumption}{Assumption}[section]
\newtheorem{definition}{Definition}
\newtheorem{remark}{Remark}%

\newtheorem{theorem}{Theorem}
\newtheorem{proposition}{Proposition}  
\setlist[enumerate]{label=(\alph*), leftmargin=*, nosep}

\newcommand{\anon}{1}

\begin{document}

\def\spacingset#1{\renewcommand{\baselinestretch}%
{#1}\small\normalsize} \spacingset{1}


\if1\anon
{
  \title{\bf Difference-in-Differences using Double Negative Controls and Graph Neural Networks for Unmeasured Network Confounding}
  \author{Zihan Zhang $^{ *}$\\
      School of Mathematics and statistics,  Liaoning University\\
      and \\ 
      Lianyan Fu  \thanks{These authors contributed equally to this
work.
      } \thanks{Corresponding author(s). E-mail(s): fulianyan@lnu.edu.cn.
            }\hspace{.2cm}\\
      School of Mathematics and statistics,  Liaoning University\\
      and \\ 
      Dehui Wang  \thanks{Corresponding author(s). E-mail(s): wangdehui@lnu.edu.cn.
      }\hspace{.2cm}\\
      School of Mathematics and statistics,  Liaoning University 
  }
  \maketitle
} \fi

\if0\anon
{
  \bigskip
  \bigskip
  \bigskip
  \begin{center}
    {\LARGE\bf }
\end{center}
  \medskip
} \fi\bigskip
\begin{abstract}
Estimating causal effects from observational network data faces dual challenges of network interference and unmeasured confounding. To address this, we propose a general Difference-in-Differences framework that integrates double negative controls (DNC) and graph neural networks (GNNs). Based on the modified parallel trends assumption and DNC, semiparametric identification of direct and indirect causal effects is established. We then propose doubly robust estimators. Specifically, an approach combining GNNs with the generalized method of moments is developed to estimate the functions of high-dimensional covariates and network structure. Furthermore, we derive the estimator's asymptotic normality under the $\psi$-network dependence and approximate neighborhood interference. Simulations show the finite-sample performance of our estimators. Finally, we apply our method to analyze the impact of China's green credit policy on corporate green innovation. 
\end{abstract}

\vfill

\newpage
\spacingset{1.8}

\section{Introduction}\label{sec-intro}
Estimating causal effects under network interference has become increasingly important in various fields. For example, scholars have explored the diffusion of microcredit within households \citep{he2024measuring}, as well as the effect of anti-conflict interventions on adolescent social norms \citep{paluck2016changing}. In the field of public health, researchers have not only examined the transmission effects of infectious diseases \citep{halloran1995causal,morozova2018risk}, but also investigated how health behaviors ``contagion'' among populations \citep{christakis2013social}. 
\par
When individuals form connections through a network, identification and estimation of causal effects have been challenging for two reasons. The first issue is that the conventional potential outcome framework \citep{rubin1980randomization}, which relies on the Stable Unit Treatment Value Assumption (SUTVA), is inappropriate. A large body of literature has focused on randomized controlled trials \citep[e.g.,][]{toulis2013estimation,athey2018exact,li2022random}. A common method to relax the SUTVA is to suppose a low-dimensional function, which is called the effective treatment \citep{manski2013identification} or exposure mapping \citep{10.1214/16-AOAS1005}. This function serves as a sufficient statistic for spillover effects, meaning that others' interventions influence an individual's outcome only through it. The exposure mapping is useful for summarizing potentially complex spillover effects, yet there is an inherent challenge in determining the ``right'' functional form. Consequently, several recent studies have explored the conditions under which meaningful causal parameters can be estimated even when interference is unknown \citep{savje2021average,leung2022,savje2024causal,hoshino2024causal}. 
\par
Beyond randomized controlled trials, a growing number of studies have considered observational causal inference in settings with network interference \citep{van2014causal,forastiere2021identification,tchetgen2021auto,leunggnn,leung2023network,ogburn2024causal}. These approaches typically rely on the unconfoundedness assumption, often requiring conditioning on the covariates of an individual and their neighbors \citep{Liu2019}, or on the covariates of all individuals in the network \citep{leunggnn}. However, a key limitation of these methods is their reliance on the assumption of no unmeasured network confounding. This assumption is frequently violated in observational settings, complicating the identification and estimation of causal effects. Such unmeasured confounding generally arises from two sources: the homophily bias (connections driven by unobserved characteristics) and the contextual confounding (peers sharing unobserved contextual factors) \citep{manski1993identification,vanderweele2013social}. 
\par
The two aforementioned issues can be summarized as the problem of unmeasured network confounding in observational studies. In this paper, we employ the double negative controls (DNC) and graph neural networks (GNNs) to address this challenge. The DNC specifies a negative control outcome---an outcome variable known not to be causally affected by the treatment---and a negative control exposure, which is a treatment variable that does not causally affect the outcome \citep{lipsitch2010negative}. In recent years, a growing body of research has used DNC as proxy variables to identify causal effects in the presence of unmeasured confounding, an approach known as proximal causal inference \citep[e.g.,][]{miao2018identifying,shi2020multiply,cui}. These methods all rely on the SUTVA and do not involve network data. Concurrently, we introduce GNNs to address the high-dimensional estimation challenges posed by network confounding in our DNC framework. \cite{leunggnn} has demonstrated that GNNs are particularly well-suited for network-structured data, capable of capturing one's own high-dimensional covariates, neighbors' high-dimensional covariates, and their complex network relationships, thereby enabling effective estimation. \cite{xu} proposed a DID framework for network interference, but the identification strategy relies on the selection on observables assumption and thus cannot address the issue of unmeasured network confounding.
\par
Our contribution is to propose a general Difference-in-Differences (DID) framework that integrates DNC and GNNs. Based on our framework, researchers are able to identify and estimate the direct and indirect causal effects of the treatment in settings with unmeasured network confounding. To achieve this, our identification strategy addresses the limitations of the standard parallel trends assumption by instead positing a latent parallel trends (cf. Assumption 2.1). Furthermore, this assumption is weaker than the network version of unconfoundedness \citep[e.g.,][]{leunggnn,egami} and forms the basis for our DNC approach. The nonparametric identification of the effects of interest is derived by leveraging DNC that are associated with unmeasured confounders. Specifically, we incorporate DNC via either an outcome confounding bridge function or a treatment confounding bridge function, which are the network-adapted versions of those studied in \cite{cui}. Identification is achieved as long as at least one of these two bridge functions satisfies the identification assumptions. Moreover, the semiparametric proximal causal inference framework permits nonparametric estimation of these quantities.
\par
Building on this identification strategy, we propose the doubly robust DID estimators. \cite{drdid} proposed doubly robust DID estimators for independent data, whereas our method accommodates settings with network confounding. Combining the generalized method of moments (GMM) \citep{hansen1982large} and GNNs, we develop an approach to handle the high-dimensional nonparametric terms within the bridge functions. Subsequently, the asymptotic normality of this estimator is established under assumptions of $\psi$-network dependence \citep{kojevnikov2021limit} and approximate neighborhood interference (ANI) \citep{leung2022}.  Finally, simulations demonstrate the finite-sample performance of the estimator, and we apply the method to evaluate the impact of the green credit policy on green innovation.
\par
The rest of the article is organized as follows. Section~\ref{sec2} details our setup, identification challenges, and our identification strategy using DNC. Section~\ref{sec3} presents the GNN-based estimation and inference. Section~\ref{sec4} evaluates our method via simulations. Section~\ref{sec5} provides an empirical application, and Section~\ref{sec6} concludes.
\par

\section{Setup and Identification}\label{sec2}
\subsection{Setup }
Consider a set of units $N_{n}$ = $\{ 1, 2, \dots, n \}$. The units form an undirected network represented by the $n$ $\times$ $n$ symmetric adjacency matrix $\mathbf{A}$ = $(A_{ij})_{i,j\in N_{n}}$, where $A_{ij}$ $\in$ $\{ 0, 1 \}$ indicates whether or not $i$ and $j$ are connected. We assume that there are no self-links so that $A_{ii}$ = 0 for all $i$ $\in$ $N_{n}$. Define $\mathcal{A}_n = \{0,1\}^{n \times n}$ as the space of all binary adjacency matrices. Let $Y_{it}$ $\in$ $\mathbb{R}$ and $D_{it}$ $\in$ $\{ 0, 1 \}$ denote the observed outcome and treatment status, respectively, for individual $i$ at time $t$, where $t$ $\in$ $\{0,1\}$. Because individuals are exposed to treatment only at $t$ = 1, we have $D_{i0}$ = 0 for all $i$. To reduce notation, we define $D_{i}$ $\equiv$ $D_{i1}$. Denote the $n$-dimensional vector of realized treatments as $\mathbf{D}$ = $(D_{i})_{i \in N_{n}}$, with the support $\mathcal{D}_{n}$ = $\{0,1\}^{n}$. For each $\mathbf{d}$ $\in$ $\mathcal{D}_{n}$, let $Y_{it}(\mathbf{d})$ denote unit $i$'s potential outcome at time $t$ under treatment assignment $\mathbf{D}$ = $\mathbf{d}$. By construction, we have $Y_{it}$ = $Y_{it}(\mathbf{D})$.  Denoting $\mathbf{d_{-i}}$ = $(d_{k})_{k\ne i}$, we write the potential outcome of unit $i$ at time $t$ as $Y_{it}(d_{i}, \mathbf{d_{-i}})$, given $D_{i}$ = $d_{i}$ and $\mathbf{D_{-i}}$ = $\mathbf{d_{-i}}$. 
\par
Because only one realization from $(Y_{it}(\mathbf{d}))_{\mathbf{d} \in \mathcal{D}_{n}}$ is observable for each unit, it is generally impossible to define identifiable causal estimands without introducing restrictions \citep{hoshino2024causal}. Accordingly, we define a pre-specified function $G$: $N_{n}$ $\times$ $\{ 0, 1 \}^{n-1}$ $\times$ $\mathcal{A}_{n}$ $\to$ $\mathcal{G}$, where $\mathcal{G}$ $\subset$ $\mathbb{R}^{dim(G)}$ is a set that does not depend on $i$ and $n$, and $dim(G)$ is a fixed positive integer. The exposure mapping $G$ defines the exposure realization for unit $i$ as $G_{i} = G(i, \mathbf{D_{-i}}, \mathbf{A})$, which maps the treatment statuses of other units ($\mathbf{D_{-i}}$) and the network structure ($\mathbf{A}$) to a specific exposure value. Following \cite{savje2024causal}, we employ $G$ to define causal effects without necessarily assuming it to fully capture the complete causal structure. For example, $G_{i}$ = $\sum_{j=1}^{n}A_{ij}D_{j}$. If $G$ is correctly specified, it implies that individual i's outcome depends only on its own treatment status and the number of treated neighbors. However, even under misspecification, $G$ remains a useful construct for defining causal estimands of interest, such as spillover effects arising from variation in the number of treated neighbors. We focus on two types of estimands defined by exposure mappings. The first estimand is the average direct effect on the treated (ADT) at exposure level $g \in \mathcal{G}$, defined as:
\begin{equation}
\tau_{ADT}(g)=\frac{1}{n}\sum_{i \in N_{n}}
E[Y_{i1}(1,\mathbf{D_{-i}})-Y_{i1}(0,\mathbf{D_{-i}})\mid G_{i}=g, D_{i}=1].
\end{equation}
For instance, if $G_{i}$ is defined as the number of treated neighbors ($G_{i} = \sum_{j=1}^{n}A_{ij}D_{j}$), then the $\tau_{ADT}$(3) represents the conditional average difference in potential outcomes between being treated and untreated for individuals who are treated and have exactly three treated neighbors.
\par
We now examine the average indirect effect on the treated (AIT) estimands. Let $\ell_{\mathbf{A}}(i,j)$ denote the path distance between units $i$ and $j$, defined as the length of the shortest path connecting them. By convention, we set $\ell_{\mathbf{A}}(i,j)$ = $\infty$ when no path exists between $i$ and $j$ in $\mathbf{A}$ and 0 if $i$ = $j$. Suppose that each $G_{i}$ depends only on unit $j$'s such that 1 $\le$ $\ell_{\mathbf{A}}(i,j)$ $\le$ $K$ with some constant $K$ $\ge$ 1  (see Assumption 4.4). Based on this, we define the interference graph $\mathbf{E}$ = $(E_{ij})_{i,j\in N_{n}}$, where $E_{ij}$ = $\mathbf{1}\{ 1 \le \ell_{\mathbf{A}}(i,j) \le K \}$. For each $i$ $\in$ $S_{n}$, let interference set $\mathcal{E} _{i}$ = $\{ j \in N_{n}: E_{ij} = 1 \}$. The AIT estimand is given by
\begin{equation}
\tau_{AIT}=\frac{1}{n}\sum_{i \in N_{n}}
E\left[\sum_{j \in \mathcal{E} _{i}}\left(Y_{j1}(D_{i}=1,\mathbf{D_{-i}})-Y_{j1}(D_{i}=0,\mathbf{D_{-i}})\right)\Bigg | D_{i}=1\right].
\end{equation}
For example, when $K$ = 2, the $\tau_{AIT}$ measures the average indirect causal effect of a treated unit's intervention status on its direct neighbors (distance 1) and neighbors of neighbors (distance 2) through network paths. Units' potential outcome may not share a common expectation, so  the causal estimand is explicitly written as  the network average of individual level causal effects. Due to space limitations, our discussion primarily focuses on ADT, while a detailed discussion of AIT is provided in Appendix C.
\subsection {Identification challenge}
The DID is one of the most popular methods in the social sciences for estimating causal effects in observational studies \citep{roth2023s}. The application of the DID to identify the effect of interest under cross-unit interference presents two main challenges. The first challenge is that when SUTVA is relaxed, the conditional parallel trends assumption does not suffice to identify the causal effect. A natural approach to adapting the conditional parallel trends assumption for identifying the ADT under cross-unit interference can be expressed as:
\begin{equation}
\begin{split}
&\quad\frac{1}{n}\sum_{i \in N_{n}}
E[Y_{i1}(0,\mathbf{D_{-i}})-Y_{i0}(0,\mathbf{D_{-i}})\mid G_{i}=g,D_{i}=1,  \mathbf{X}, \mathbf{A}]\\
&=\frac{1}{n}\sum_{i \in N_{n}}
E[Y_{i1}(0,\mathbf{D_{-i}})-Y_{i0}(0,\mathbf{D_{-i}})\mid G_{i}=g, D_{i}=0, \mathbf{X},\mathbf{A}], 
\end{split}
\end{equation}
where $\mathbf{X}$ = $(X_{i})_{i=1}^{n}$ represents  the matrix of all units' observed  covariates, and $X_{i}$ $\in$ $\mathbb{R}^{d}$. However, due to the presence of unmeasured confounding $\mathbf{U}$ = $(U_{i})_{i=1}^{n}$, this assumption generally fails to hold. This assumption is similar to Assumption 3 in \cite{xu}. Figure 1 shows a causal-directed acyclic graph (DAG). Without loss of generality, we assume there are only two samples, $\Delta Y_{i} = Y_{i1} - Y_{i0}$ for $i \in \{1, 2\}$. If $D_{1} = 1$ and $D_{2} = 0$, then $D_{1} \to \Delta Y_{1}$ and $D_{1} \to \Delta Y_{2}$ denote the ADT and the AIT, respectively. When equation (3) holds, there exists an unblocked back-door path $D_{1}$ $\gets$ $\mathbf{U}$ $\to$ $\Delta Y_{1}$.  Consequently, even in the simple setting of dyadic data, identification of the ADT requires additional assumptions. As the identification challenges for the AIT are analogous, we omit a separate discussion here.
\begin{figure}[h!]
\centering
\includegraphics[width=0.3\textwidth]{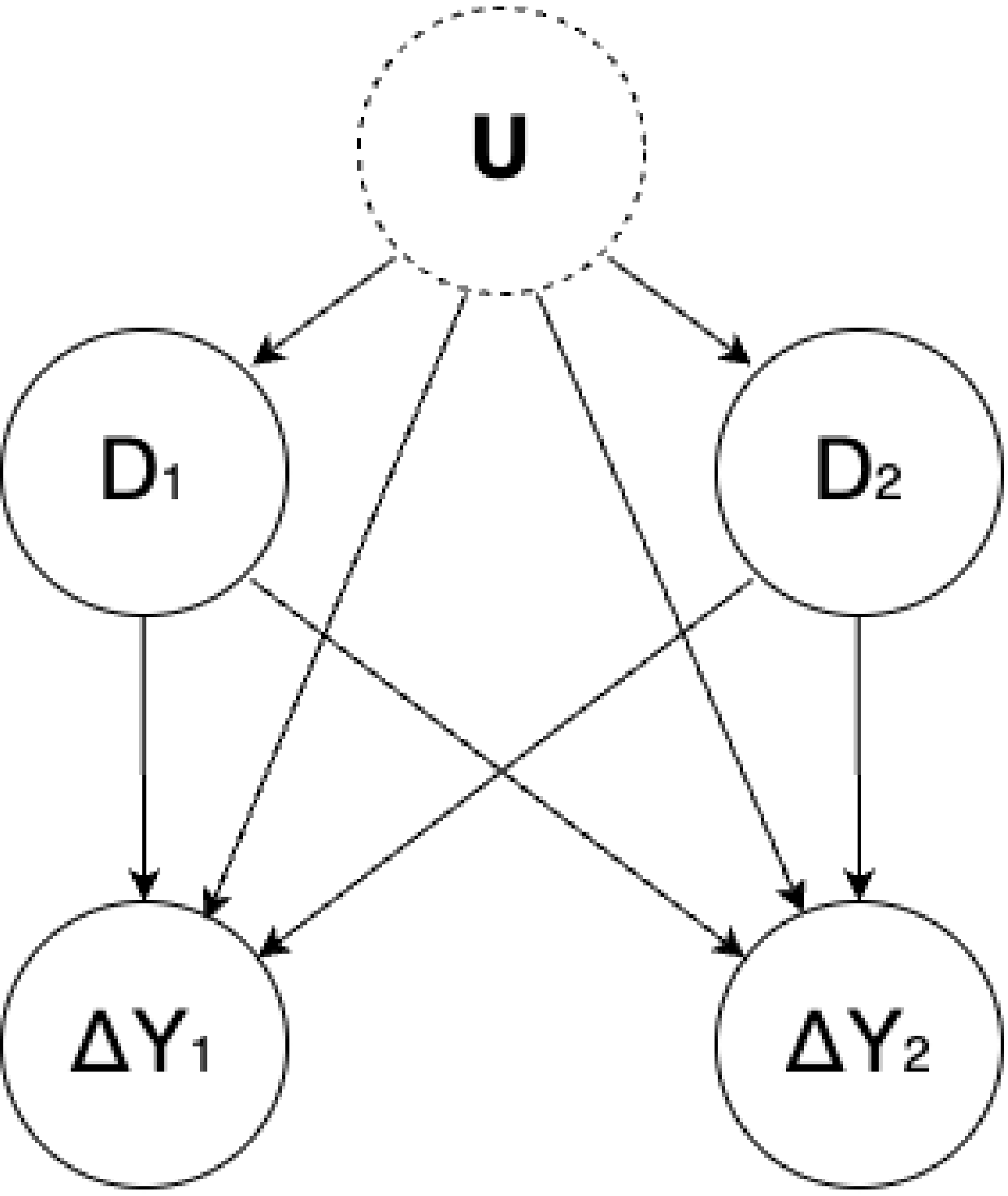}
\caption{\label{fig-first}DAGs for dyadic data in the presence of unmeasured confounding. Solid (dotted) nodes denote observed (unobserved) variables, respectively, although the observed covariates $\mathbf{X}$ are suppressed for simplicity.}
\end{figure}\subsection { Identification with double negative controls}
In this section, we consider an alternative identification strategy that leverages auxiliary variables known as negative controls to address unmeasured confounding. While measuring all confounding factors would ensure the standard conditional parallel trends assumption (3) holds, this is often infeasible. We therefore posit the following latent parallel trends assumption.
\begin{assumption}[Latent parallel trends for ADT].
\begin{equation}
\begin{split}
\nonumber
&\quad\frac{1}{n}\sum_{i \in N_{n}}
E[Y_{i1}(0,\mathbf{D_{-i}})-Y_{i0}(0,\mathbf{D_{-i}})\mid G_{i}=g, D_{i}=1, \mathbf{U}, \mathbf{X}, \mathbf{A}]\\
&=\frac{1}{n}\sum_{i \in N_{n}}
E[Y_{i1}(0,\mathbf{D_{-i}})-Y_{i0}(0,\mathbf{D_{-i}})\mid G_{i}=g, D_{i}=0, \mathbf{U}, \mathbf{X},\mathbf{A}]. 
\end{split}
\end{equation}
\end{assumption}
This is similar to the latent ignorability of \cite{egami}. Assumption 2.1 states that $\mathbf{U}$ and $\mathbf{X}$ suffice to account for confounding,  whereas $\mathbf{X}$ alone may not. This assumption is often plausible as there is no direct restriction on the nature of the latent characteristic $\mathbf{U}$. Additionally, we impose the no anticipation assumption, stipulating that units do not alter their behavior in period $0$ in expectation of future treatment assignment.
\begin{assumption}[No anticipation].
\begin{equation}
\nonumber
Y_{i0}(d, \mathbf{d_{-i}}) = Y_{i0}(0, \mathbf{0}) \quad \text{for all } d \in \{0,1\}, \mathbf{d_{-i}} \in \{0,1\}^{n-1}.
\end{equation}
\end{assumption}
This assumption is widely adopted in the studies on the DID models with interference \citep[e.g.,][]{butts2021difference,deuchert2019direct}. However, Assumptions 2.1-2.2 are not sufficient for identification due to the unmeasured variables $\mathbf{U}$. For identification under the unobserved confounding, we incorporate two observed auxiliary variables, NCE $Z$, and NCO $W$, which satisfy the following conditions.
\begin{assumption}[Negative controls].
\\
(a) Negative control outcome (NCO): For all $i$ $\in$ $N_{n}$, $W_{i}$ satisfy
\begin{equation}
\nonumber
 W_{i}\perp D_{i} \mid G_{i},\mathbf{U}, \mathbf{X}, \mathbf{A}.
\end{equation}
(b) Negative control exposure (NCE): For all $i$ $\in$ $N_{n}$, $Z_{i}$ satisfy
\begin{equation}
\begin{split}
\nonumber
&Z_{i}\perp \Delta Y_{i} \mid D_{i},G_{i},\mathbf{U}, \mathbf{X}, \mathbf{A},\quad \text{and}
\\&Z_{i}\perp  W_{i} \mid D_{i},G_{i},\mathbf{U}, \mathbf{X}, \mathbf{A}.
\end{split}
\end{equation}
\end{assumption}
Assumption 2.3(a) defines the NCO property, which states that $W_{i}$ is conditionally independent of the treatment $D_{i}$ given the unmeasured confounders $\mathbf{U}$, observed covariates $\mathbf{X}$, network structure $\mathbf{A}$, and exposure mapping $G_{i}$. Assumption 2.3(b) defines the NCE property, which states that $Z_{i}$ is conditionally independent of both the outcome $\Delta Y_{i}$ and the NCO $W_{i}$, given $\mathbf{U}, \mathbf{X}, \mathbf{A}, G_{i}$, and $D_{i}$.
\par
Identifying the ADT using NCO W and NCE Z hinges on a key assumption within the proximal causal inference framework \citep{miao2018identifying}: the existence of a confounding bridge function, as specified in Assumption 2.4.
\begin{assumption}[Outcome confounding bridge function]. 
There exists a function $h_1$($W_{i}$, $G_i$, $\mathbf{X}$, $\mathbf{A}$), such that for all $g$ $\in$ $\mathcal{G}$, and all $i$ $\in$ $N_{n}$,
\begin{equation}
\begin{split}
E[\Delta Y_{i}\mid G_{i}=g, D_{i}=0, \mathbf{U}, \mathbf{X}, \mathbf{A}]
=
E[h_1( W_{i}, \mathbf{X}, \mathbf{A})\mid G_{i}=g,D_{i}=0,  \mathbf{U}, \mathbf{X},\mathbf{A}]. 
\end{split}
\end{equation}
\end{assumption}
Assumption 2.4 states that the confounding effect of $\mathbf{U}$ on outcome $\Delta Y_{i}$ is equal to the confounding effect of $\mathbf{U}$ on outcome confounding bridge function $h_1$($W_{i}$, $\mathbf{X}$, $\mathbf{A}$),  a transformation of $W_{i}$. Since the first term in the ADT contrast is point-identified under the consistency assumption alone, we only requires invoking the confounding bridge function for the counterfactual outcome under the control  group. Equation (4) is formally a  Fredholm integral equation of the first kind \citep{kress1989linear}. The existence of a solution to this equation, as detailed in Appendix A.1 of \cite{egami}, demands that $W$ be sufficiently informative for $U$, in addition to a set of regularity conditions. 
\par
\begin{remark}[Feasibility of Estimation with GNNs]
Without network confounding, Equation (4) can be expressed as:
\begin{equation}
\begin{split}
\nonumber
E[\Delta Y_{i}\mid G_{i}=g, D_{i}=0, U_i, X_i]
=
E[h_1( W_{i}, \mathbf{X}, \mathbf{A})\mid G_{i}=g,D_{i}=0,  U_i, X_i]. 
\end{split}
\end{equation}
This equation takes the form of a Fredholm integral equation of the first kind \citep{kress1989linear}. It admits a solution under certain regularity conditions; alternatively, the bridge function can be estimated using nonparametric methods \citep[e.g.,][]{cui,egami}. 
\par
However, in our setting with unmeasured network confounding, the bridge function $h_1(W_i, \mathbf{X}, \mathbf{A})$ depends on the high-dimensional global covariates $\mathbf{X}$ and the network structure $\mathbf{A}$. Estimating such a fully flexible function from a single network observation is generally infeasible due to the curse of dimensionality. We address this challenge by parameterizing $h_1$ with GNNs. GNNs impose a structural constraint of permutation invariance, which allows the model to learn a common functional form from local variations across the graph. Furthermore, under the assumption of $\psi$-weak dependence (Assumption 3.6), the spatial averages over the single large network converge to the superpopulation expectations, rendering the estimation feasible. For a detailed discussion of permutation invariance, see Section 3.3 of \cite{leunggnn}.
\end{remark}
\par
To achieve identification under unobserved confounding, we next employ the NCE $Z$. 
\begin{assumption}[Negative control relevance]. 
 For any square integrable function $f$ and any $g$, $\mathbf{x}$ and $\mathbf{A}$, if E(f($W_{i}$) $\mid$ $G_{i}$ = $g$, $D_{i}$ = $0$, $Z_{i}$ =$z$, $\mathbf{X}$ = $\mathbf{x}$, $\mathbf{A}$) = 0 for almost all $z$, then f($W_{i}$) = 0 almost surely.
\end{assumption}
This assumption states that $Z$ contains sufficient information about $W$, which is crucial for identifying the outcome confounding bridge function $h_1$. It is a well-known technical condition in the study of sufficiency in statistical inference, called the completeness condition. In practice, selecting NCO and NCE that satisfy Assumption 2.5 is crucial for effect estimation. \cite{egami} discuss the impacts on estimation when the assumption is violated. To satisfy the completeness condition for categorical variables, the number of categories in NCE needs to be at least as large as the number of categories in NCO. For continuous variables, the number of NCEs needs to be at least as large as the number of NCOs.

\par
The selection of valid negative controls in practice is crucial. Figure 2 illustrates an example of how to choose negative controls that satisfy Assumption 2.2. We consider a case with four samples where the effect of interest is $D_{1} \to \Delta Y_{1}$. In this DAG framework, the treatment assigned to each unit exerts direct effects only on itself and its immediate neighbors. $D_{2}$ and the two variables $\{D_{3}, D_{4}\}$ satisfy the NCO and NCE conditions, respectively. This negative control selection approach is applicable when there are no direct causal relationships between individual treatments, and requires that the number of first-order neighbors (distance $s$ = 1) is substantially smaller than the number of individuals at distance $s$ $\ge$ 2. 
\begin{figure}[h!]
\centering 
\includegraphics[width=0.5\textwidth, trim={0cm 6.5cm 0cm 6.5cm}, clip]{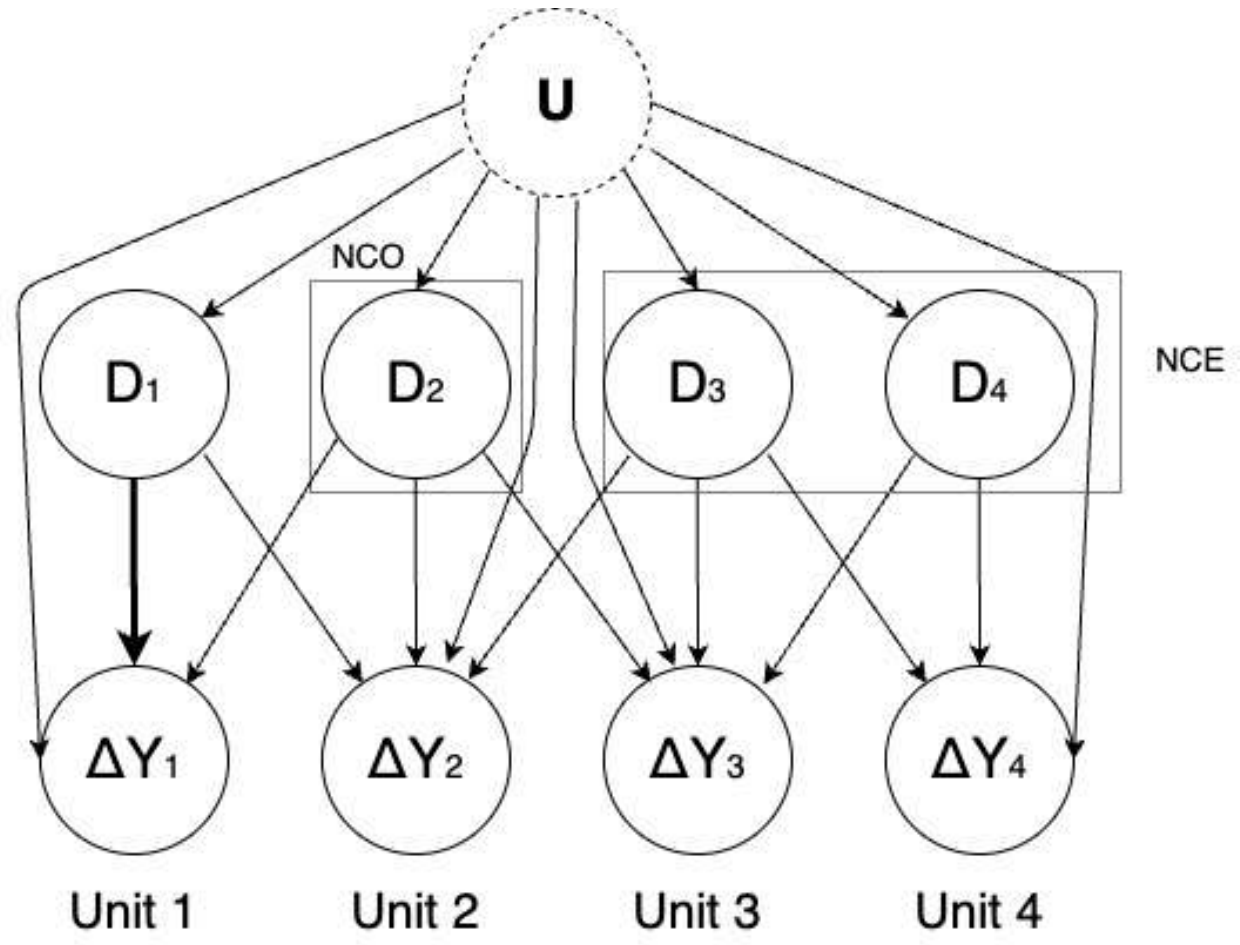}\caption{\label{fig-first}DAGs for network data in the presence of unmeasured confounding. The thick arrows from $D_{1}$ to $\Delta Y_{1}$  indicates the causal effect of interest. Solid (dotted) nodes denote observed (unobserved) variables, respectively, although the observed covariates $\mathbf{X}$ are suppressed for simplicity.}
\end{figure}%
\par
Under the stated assumptions, we establish the nonparametric identification of the ADT.
\begin{theorem}\label{theorem1}
Under Assumption 2.1-2.5, the confounding bridge function is identified as the unique solution to the following equation:
\begin{equation}
\begin{split}
E[\Delta Y_{i}\mid G_{i}=g, D_{i}=0, Z_{i}, \mathbf{X}, \mathbf{A}]
=
E[h_1(W_{i}, \mathbf{X}, \mathbf{A})\mid G_{i}=g, D_{i}=0, Z_{i}, \mathbf{X}, \mathbf{A}], 
\end{split}
\end{equation}
and the ADT(g) is identified by
\begin{equation}
\begin{split}
\nonumber
\tau_{ADT}(g)=\frac{1}{n}\sum_{i \in N_{n}}
E[\Delta Y_{i}
-h_1(W_{i}, \mathbf{X}, \mathbf{A})
\mid G_{i}=g, D_{i}=1].
\end{split}
\end{equation}
\end{theorem}
Notably, we identify the ADT without imposing any parametric restriction on the confounding bridge function, and the proof of Theorem~\ref{theorem1} can be find in Appendix A.1. 
\par

\subsection {Doubly robust difference-in-differences estimands}
In this section, we examine an alternative identification approach \citep{cui} and subsequently derive the doubly robust DID estimands.
\begin{assumption}[ Treatment confounding bridge function]. 
There exists a function $q_1$($Z_{i}$, $\mathbf{X}$, $\mathbf{A}$), such that for all $g$ $\in$ $\mathcal{G}$, and all $i$ $\in$ $N_{n}$,
\begin{equation}
\begin{split}
\nonumber
\frac{E[\mathbf{1}_i(g,1)\mid \mathbf{U}, \mathbf{X}, \mathbf{A}]}
{E[\mathbf{1}_i(g,0)\mid  \mathbf{U}, \mathbf{X}, \mathbf{A}]}
=
E[q_1( Z_{i}, \mathbf{X}, \mathbf{A})\mid G_{i}=g,D_{i}=0,  \mathbf{U}, \mathbf{X},\mathbf{A}],
\end{split}
\end{equation}
where $\mathbf{1}_i(g,d)$ = $\mathbf{1}\{G_i=g, D_i=d\}$ for $g$ $\in$ $\mathcal{G}$ and $d$ $\in$ $\{0,1\}$.
\end{assumption}
\begin{assumption}[Negative control relevance]. 
 For any square integrable function $f$, $g$, $\mathbf{x}$ and $\mathbf{A}$, if E(f($Z_{i}$) $\mid$ $G_{i}$ = $g$, $D_{i}$ = $0$, $W_{i}$ =$w$, $\mathbf{X}$ = $\mathbf{x}$, $\mathbf{A}$) = 0 for almost all $w$, then f($Z_{i}$) = 0 almost surely.
\end{assumption}
Analogous to Assumptions 2.4 and 2.5, Assumptions 2.6 and 2.7 allow us to establish another non-parametric identification of the ADT.
\begin{theorem}\label{theorem2}
Under Assumption 2.1-2.5, the confounding bridge function is identified as the unique solution to the following equation:
\begin{equation}
\begin{split}
\frac{E[\mathbf{1}_i(g,1)\mid W_i, \mathbf{X}, \mathbf{A}]}
{E[\mathbf{1}_i(g,0)\mid W_i, \mathbf{X}, \mathbf{A}]}
=
E[q_1( Z_{i}, \mathbf{X}, \mathbf{A})\mid G_{i}=g,D_{i}=0, W_i, \mathbf{X},\mathbf{A}],
\end{split}
\end{equation}
and the ADT(g) is identified by
\begin{equation}
\begin{split}
\nonumber
\tau_{ADT}(g)=\frac{1}{n}\sum_{i \in N_{n}}\left( 
E[\frac{\mathbf{1}_i(g,1)\Delta Y_{i}}
{E(\mathbf{1}_i(g,1))}]
-
E[\frac{q_1(Z_{i}, \mathbf{X}, \mathbf{A})\mathbf{1}_i(g,0)\Delta Y_{i}}
{E(\mathbf{1}_i(g,1))}] \right).
\end{split}
\end{equation}
\end{theorem}
Theorem~\ref{theorem2} provides a new identification result for the ADT, and its proof can be found in Appendix A.2.
\par
To estimate causal effects, a common approach involves specifying a parametric or semi-parametric function for the confounding bridge function $h_1(W_{i}, \mathbf{X}, \mathbf{A}; \gamma_1)$ with $\gamma_1$, or $q_1(Z_{i}, \mathbf{X}, \mathbf{A}; \gamma_2)$ with $\gamma_2$ \citep{miao2018identifying,egami,cui}. Then, based on the conditional moment restriction in Theorem 1 (for $h_1$) or Theorem 2 (for $q_1$), one can treat the conditioning set as a high-dimensional vector function and estimate $\gamma_1$ or $\gamma_2$ via GMM method. However, this method relies on the correct specification of the outcome confounding bridge function and the treatment confounding bridge function. If either the function $h_1$ or $q_1$ are misspecified, the plug-in estimator may be severely biased. 
\par
To address this challenge, we propose the doubly robust DID estimand tailored for unmeasured network confounding. Here, the doubly robust property implies that  the resulting estimand identifies the the causal effect of interest, as long as at least one of confounding bridge functions is correctly specified. Let $h^*_1(W_{i}, \mathbf{X}, \mathbf{A})$ and $q^*_1(Z_{i}, \mathbf{X}, \mathbf{A})$ be arbitrary models for the true, unknown function $h_1(W_{i}, \mathbf{X}, \mathbf{A})$ and $q_1(Z_{i}, \mathbf{X}, \mathbf{A})$ for $i$ $\in$ $N_n$. In order to describe our proposed doubly robust approach, consider the following two models:
\begin{enumerate}
  \item Model $\mathcal{M}_1$, in which $h^*_1(W_{i}, \mathbf{X}, \mathbf{A})$ = $h_1(W_{i}, \mathbf{X}, \mathbf{A})$ and Assumptions 2.1-2.5 hold.
  \item Model $\mathcal{M}_2$, in which $q^*_1(Z_{i}, \mathbf{X}, \mathbf{A})$ = $q_1(Z_{i}, \mathbf{X}, \mathbf{A})$ and Assumptions 2.1-2.3, and 2.6-2.7 hold.
  \end{enumerate}
The doubly robust DID estimand can be expressed as:
\begin{equation}
\begin{split}
\nonumber
\tau^{dr}_{ADT}(g)=\frac{1}{n}\sum_{i \in N_{n}} 
E\left[\left(\frac{\mathbf{1}_i(g,1)}
{E(\mathbf{1}_i(g,1))}
-
\frac{q^*_1(Z_{i}, \mathbf{X}, \mathbf{A})\mathbf{1}_i(g,0)}
{E(\mathbf{1}_i(g,1))}\right)
\left(\Delta Y_{i}-h^*_1(W_{i}, \mathbf{X}, \mathbf{A})\right)\right] .
\end{split}
\end{equation}
The following Proposition~\ref{proposition1} shows that our proposed doubly robust estimand recovers the ADT provided that at least one of models $\mathcal{M}_1$ and $\mathcal{M}_2$ is correctly specified.
\begin{proposition}\label{proposition1}
If at least one of the models $\mathcal{M}_1$ and $\mathcal{M}_2$ is correctly specified, then $\tau^{dr}_{ADT}(g)$ = $\tau_{ADT}(g)$.
\end{proposition}
The proof can be find in Appendix A.3. 

\section{Estimation and Inference}\label{sec3}
In this section, we discuss the estimation and inference for ADT while accounting for network-dependent in observational studies. 
\subsection{Doubly robust difference-in-differences estimators using graph neural networks}
Our estimation procedure can be decomposed into two steps. In the first step, GNNs are employed to estimate the working nuisance models $h^*_1(W_{i}, \mathbf{X}, \mathbf{A})$ and $q^*_1(Z_{i}, \mathbf{X}, \mathbf{A})$, $i$ $\in$ $N_n$, $g$ $\in$ $\mathcal{G}$. In the second step, one plugs the fitted values of the working nuisance models into the sample analogue of $\tau^{dr}_{ADT}(g)$. 
\par
The standard GNNs architecture comprises $L$ layers, each containing $n$ computational units known as neurons. Each neuron, denoted as $h_i^{(l)}$ for the $i$-th neuron in layer $l$, implements a parameterized, vector-valued function. The connectivity between layers is governed by adjacency matrix $\mathbf{A}$ via a ``message-passing'' mechanism, for layers $l$ = 1, $\dots$, $L$,
\begin{equation}
\begin{split}
\nonumber
h_i^{(l)} = \Phi_{0l} \Biggl( 
    h_i^{(l-1)},\  
    \Phi_{1l} \biggl( 
        h_i^{(l-1)},\ 
        \Bigl\{ h_j^{(l-1)} : A_{ij} = 1 \Bigr\} 
    \biggr) 
\Biggr),
\end{split}
\end{equation}
where $\Phi_{0l}$, $\Phi_{1l}$ are parameterized, vector-valued functions, and $h_i^{(0)}$ = $X_i$ denotes the initial node, which contains no inherent network information. In subsequent layers, unit $i$'s node embedding is a function of its $1$-neighborhood's embeddings in the previous layer and therefore incorporates increasingly more network information as $l$ increases. 
\par
For estimating the confounding bridge functions, we utilize a strategy that combines GNNs with the GMM method. Let $\mathcal{F}_{GNN}(L)$ denote the set of all GNNs with $L$ layers ranging over all possible functions $\Phi_{0l}$, $\Phi_{1l}$ for $l$ = 1, $\dots$, $L$ within some function class. For any $f$ $\in$ $\mathcal{F}_{GNN}(L)$, let $f(i, \mathbf{X}, \mathbf{A})$ denote its $i$th component, which corresponds to $h_i^{(L)}$. To allow for flexible non-linear interactions while strictly satisfying the identification condition that bridge functions depend on unit-specific negative controls, we employ a late-fusion neural network architecture. Suppose both functions $h^*_1(W_{i}, \mathbf{X}, \mathbf{A})$ and $q^*_1(Z_{i}, \mathbf{X}, \mathbf{A})$ take the following form:
\begin{equation} 
\begin{split} 
\nonumber
h^*_1(W_i, \mathbf{X}, \mathbf{A}) = \Psi_h \left( W_i, f(i, \mathbf{X}, \mathbf{A}) \right), \ q^*_1(Z_i, \mathbf{X}, \mathbf{A}) = \Psi_q \left( Z_i, f(i, \mathbf{X}, \mathbf{A}) \right), 
\end{split} 
\end{equation}
where $\Psi_h$ and $\Psi_q$ are multilayer perceptrons (MLPs) that take the concatenation of the negative control variable and the GNN-learned embedding $f(i, \mathbf{X}, \mathbf{A})$ as input. This formulation replaces the restrictive linear specification with a general non-parametric form.
\par
Based on equation (5), a moment function is defined for each $i$ as:
\begin{equation}
\begin{split}
\nonumber
m_h(\Delta Y_i,W_i,D_i,G_i,Z_{i}, \mathbf{X}, \mathbf{A})
=(\Delta Y_{i}-h^*_1(W_{i}, \mathbf{X}, \mathbf{A}))\mathbf{1}_i(g,0)\begin{bmatrix}
    Z_i \\
    (\mathbf{A}\mathbf{X})^{\text{T}}_{i\cdot}
\end{bmatrix},
\end{split}
\end{equation}
where $(\mathbf{A}\mathbf{X})_{i\cdot} \in \mathbb{R}^d$ denotes the $i$-th row of the matrix product $\mathbf{A}\mathbf{X}$, which is the vector sum of the feature vectors of all individuals directly connected to individual $i$. Then, the estimator of $h^*_1$ is
\begin{equation}
\begin{split}
\nonumber
\hat{h}^*_1(W_{i}, \mathbf{X}, \mathbf{A})
= \operatorname*{argmin}_{h^*_1} \bar{m}^{\text{T}}(h^*_1)\Omega\bar{m}(h^*_1),
\end{split}
\end{equation}
where $\bar{m}(h^*_1)=1/n\sum_{i\in N_n}m_h(\Delta Y_i,W_i,D_i,G_i,Z_{i}, \mathbf{X}, \mathbf{A})$, and $\Omega$ is a user-specified positive-definite weight matrix.
\par
Next, $q^*_1(Z_{i}, \mathbf{X}, \mathbf{A})$ is estimated based on (6). Similarly, define a moment function for each $i$ as 
\begin{equation}
\begin{split}
\nonumber
m_q(W_i,D_i,G_i,Z_{i}, \mathbf{X}, \mathbf{A})
=(q^*_1\left(Z_{i},\mathbf{X}, \mathbf{A})\mathbf{1}_i(g,0)-\mathbf{1}_i(g,1)\right)\begin{bmatrix}
    W_i \\
    (\mathbf{A}\mathbf{X})^{\text{T}}_{i\cdot}
\end{bmatrix}.
\end{split}
\end{equation}
Then, the estimator of $q^*_1$ is
\begin{equation}
\begin{split}
\nonumber
\hat{q}^*_1(Z_{i}, \mathbf{X}, \mathbf{A})
= \operatorname*{argmin}_{q^*_1} \bar{m}^{\text{T}}(q^*_1)\Omega\bar{m}(q^*_1),
\end{split}
\end{equation}
where $\bar{m}(q^*_1)=1/n\sum_{i\in N_n}m_q(W_i,D_i,G_i,Z_{i}, \mathbf{X}, \mathbf{A})$.
\par
To obtain an efficient estimator, we use the two-step GMM to get the optimal estimator \citep{hansen1982large}. In the first step, we choose an identity matrix as $\Omega$ or some other positive-definite matrix, and compute the preliminary estimate $\hat{h}^*_1(1)$. In the second step, we compute $\hat{\Lambda}$ by
 \begin{equation}
\begin{split}
\nonumber
\hat{\Lambda}=\frac{1}{n}\sum_{i\in N_n}
(\Delta Y_{i}-\hat{h}^*_1(1))\mathbf{1}_i(g,0)\begin{bmatrix}
    Z_i \\
    (\mathbf{A}\mathbf{X})^{\text{T}}_{i\cdot}
\end{bmatrix}
(\Delta Y_{i}-\hat{h}^*_1(1))\mathbf{1}_i(g,0){\begin{bmatrix}
    Z_i \\
    (\mathbf{A}\mathbf{X})^{\text{T}}_{i\cdot}
\end{bmatrix}}^{\text{T}}.
\end{split}
\end{equation} 
Then, the final estimate is obtained using the weighting matrix $\Omega$ = $\hat{\Lambda}^{-1}$. The $q^*_1$ is estimated in a similar manner.

\par
The number of layers $L$ in a GNNs determines its receptive field, meaning each i's estimate is constructed using information exclusively from its $L$-neighborhood. As discussed in \cite{leunggnn}, the ANI assumption (see Assumption 3.4)  is analogous to the approximate sparsity assumption in lasso literature, demonstrating that selecting a small $L$ and  making estimation of the nuisance functions feasible. Finally, the doubly robust estimator of ADT is given by
\begin{equation}
\begin{split}
\nonumber
\hat{\tau}^{dr}_{ADT}(g)=\frac{1}{n}\sum_{i \in N_{n}} 
\left(\frac{\mathbf{1}_i(g,1)}
{\hat{E}(\mathbf{1}_i(g,1))}
-
\frac{\hat{q}^*_1(Z_{i}, \mathbf{X}, \mathbf{A})\mathbf{1}_i(g,0)}
{\hat{E}(\mathbf{1}_i(g,1))}\right)
\left(\Delta Y_{i}-\hat{h}^*_1(W_{i}, \mathbf{X}, \mathbf{A})\right),
\end{split}
\end{equation}
where $\hat{\mathrm{E}}[\mathbf{1}_i(g,1)]$ = $1/n\sum_{i\in N_n}\mathbf{1}_i(g,1)$, $h^*_1(W_{i}, \mathbf{X}, \mathbf{A})$ and $q^*_1(Z_{i}, \mathbf{X}, \mathbf{A})$ are constructed based on the aforementioned GNNs estimators. The following assumption imposes restrictions on the convergence rate of the GNNs estimators.
\begin{assumption}[GNNs Rates].
\\
For any $i$ $\in$ $N_n$ and $g$ $\in$ $\mathcal{G}$,suppose \\(a) $\frac{1}{n}\sum_{i\in N_n} (\hat{h}^*_1(W_{i}, \mathbf{X}, \mathbf{A})-h^*_1(W_{i}, \mathbf{X}, \mathbf{A}))^2=o_P(1)$ and  $\frac{1}{n}\sum_{i\in N_n} (\hat{q}^*_1(Z_{i}, \mathbf{X}, \mathbf{A})-q^*_1(Z_{i}, \mathbf{X}, \mathbf{A}))^2=o_P(1)$. \\(b)  $\frac{1}{n}\sum_{i\in N_n} (\hat{h}^*_1(W_{i},  \mathbf{X}, \mathbf{A})-h^*_1(W_{i},\mathbf{X}, \mathbf{A}))^2(\hat{q}^*_1(Z_{i}, \mathbf{X}, \mathbf{A})-q^*_1(Z_{i}, \mathbf{X}, \mathbf{A}))^2=o_P(n^{-1})$. \\(c) $\frac{1}{n}\sum_{i \in N_{n}}  \left[(\frac{\mathbf{1}_i(g,1)}{\mathrm{E}[\mathbf{1}_i(g,1)]}
   - \frac{\mathbf{1}_i(g,0)q^*_1(Z_{i},  \mathbf{X}, \mathbf{A})}{\mathrm{E}[\mathbf{1}_i(g,1)]})
   ( \hat{h}^*_1(W_{i},  \mathbf{X}, \mathbf{A})
-h^*_1(W_{i}, \mathbf{X}, \mathbf{A}))
   \right]=o_P(n^{-1/2})$.
\end{assumption}

These are similar with the standard conditions \citep[e.g., Assumption 3 of][]{farrell2015robust} for machine learners. \cite{leunggnn} examines the validity of this assumption for GNNs estimators and \cite{chen2024causal} provides the rate of convergence for artificial neural networks under certain regularization conditions. Assumption 3.1 (a) is a mild consistency requirement for the GNNs estimators. Assumption 3.1 (b)  requires an explicit rate for the product of errors. Thus, if one function is relatively easy to estimate, Assumption 3.1 (b) can still be satisfied even when the other does not converge at the $n^{-1/4}$. Assumption 3.1 (c) is similar with Assumption 3 (c) of \cite{farrell2015robust}, it constrains the convergence rate of the product between the GNNs estimators' error and some terms whose expectations are zero.

\subsection{Inference}
We next examine the convergence properties of the ADT estimator and impose the following conditions.
\begin{assumption}[Bounded outcome].
\\
There exists a constant $\bar{Y}$ such that $|Y_{it}(d, \mathbf{d_{-i}})|$ $\le$ $\bar{Y}$ $\le$ $\infty$ for all $i$ $\in$ $N_n$, $t$ $\in$ $\{0,1\}$, $d$ $\in$ $\{0,1\}$, $\mathbf{d_{-i}}$ $\in$ $\{0,1\}^{n-1}$.
\end{assumption}
\begin{assumption}[Overlap].
\\
There exist constants $C_1$, $C_2$ $\in$ (0, 1) such that $P(\mathbf{1}_i(g,d) = 1)$ $\in$ $(C_1, C_2)$, $E[\mathbf{1}_i(g,d)|W_i, \mathbf{X}, \mathbf{A}]$ $\in$ $(C_1, C_2)$ for all $d$ $\in$ $\{0,1\}$ and $g$ $\in$ $\mathcal{G}$.
\end{assumption}
Assumption 3.2 is standard condition and can be generalized to uniformly bounded moments. Assumption 3.3 ensures that the set used for estimation is non-empty. For a nonnegative integer $s$ $\ge$ 0,  let $N_{\mathbf{A}}(i,s)$ = $\{j \in N_n: \ell_{\mathbf{A}}(i,j) \le s \}$ denote the set of units within a distance $s$ from unit $i$, referred to as the $s$-neighborhood of unit $i$ (Note that $i$ $\in$ $N_{\mathbf{A}}(i,s)$). Define $\mathbf{d}_{N_{\mathbf{A}}(i,s)}$ = ($d_j: j \in N_{\mathbf{A}}(i,s)$) and $\mathbf{A}_{N_{\mathbf{A}}(i,s)}$ = ($A_{kl}: k,l \in N_{\mathbf{A}}(i,s)$),  respectively the subvector of $\mathbf{d}$ and subnetwork of $\mathbf{A}$ on $N_{\mathbf{A}}(i,s)$. Additionally, let $N^c_{\mathbf{A}}(i,s)$ = $N_n \texttt{\textbackslash} N_{\mathbf{A}}(i,s)$ denote the set of units who are more than distance $s$ away from $i$.
\begin{assumption}[Relevance limitations].
\\
There exists a known positive integer $K$ $\in$ $\mathbb{N}$ such that: (a) for all $i$ $\in$ $N_n$, $\mathbf{A}$, $\mathbf{A}'$ $\in$ $\mathcal{A}_{n}$, and $\mathbf{d}$, $\mathbf{d}'$ $\in$ $\mathcal{D}_{n}$,
\begin{equation}
\begin{split}
\nonumber
&\quad \quad N_{\mathbf{A}}(i,K)=N_{\mathbf{A}'}(i,K), 
\mathbf{A}_{N_{\mathbf{A}}(i,K)}=\mathbf{A'}_{N_{\mathbf{A'}}(i,K)},\ and \ \  \mathbf{d}_{N_{\mathbf{A}}(i,s)} = \mathbf{d}'_{N_{\mathbf{A'}}(i,s)}
\\&\Longrightarrow G(i,\mathbf{d}_{-i},\mathbf{A})=G(i,\mathbf{d}_{-i}',\mathbf{A}');
\end{split}
\end{equation}
(b) for all $i$ $\in$ $N_n$, Cov($D_i$, $\mathbf{D}_{N^c_{\mathbf{A}}(i,K)}$) = 0. (c) $\{(G_i, D_i)\}_{i \in N_n}$ are identically distributed across $i$ $\in$ $N_n$ for estimating the parameters conditioned on $(G_i,D_i)$ = $(g, d)$.
\end{assumption}
The Assumption 3.4 (a) states that the exposure mapping of each unit depends only on the unit's own K-neighborhood. This is a weak restriction on $G$ satisfied by most exposure mappings of interest in the literature \citep{leung2022}. And Assumption 3.4(b) implies that $D_i$ is unrelated to the treatment of individuals beyond its $K$-neighborhood.
\par
Below we introduce the ANI assumption. Let $\mathbf{D'}$ be an independent copy of $\mathbf{D}$, define $\mathbf{D}_i^{(s)}$ = $(\mathbf{D}_{N_{\mathbf{A}}(i,s)}, \mathbf{D}'_{N^c_{\mathbf{A}}(i,s)})$ obtained by concatenating the
subvector of $\mathbf{D}$ on $N_{\mathbf{A}}(i,s)$ and that of $\mathbf{D}'$ on $N_n\texttt{\textbackslash}N_{\mathbf{A}}(i,s)$. Finally, let
\begin{equation}
\begin{split}
\nonumber
\theta_{n,s}^{ADT}=\max_{i\in N_n}E[|\Delta Y_{i}(\mathbf{D})-\Delta Y_{i}(\mathbf{D}_i^{(s)})| \mid \mathbf{Z}, \mathbf{X}, \mathbf{A}].
\end{split}
\end{equation}
\begin{assumption}[ANI for ADT].
\\
\begin{equation}
\begin{split}
\nonumber
\sup_{n\in \mathbb{N}}\theta_{n,s}^{ADT}\to 0 \quad \text{as}\quad s\to \infty.
\end{split}
\end{equation}
\end{assumption}
This requires interference from distant alters to be negligible for large distances. The ANI condition is substantially less restrictive than the conventional clustered interference assumption, which imposes the stringent requirement that $\theta_{n,s}^{ADT}$ =0 for some finite cutoff distance $s$.
\par
Given our analysis involves a network of arbitrarily interconnected units, the independently and identically distributed assumption becomes untenable. To derive the asymptotic distribution, we employ the central limit theorem (CLT) for $\Psi$-network dependence established by \cite{kojevnikov2021limit}. 
\begin{definition}[$\Psi$-network dependence]\label{D1}.
\\
For any $H$, $H'$ $\subseteq$ $N_n$, define $\ell_{\mathbf{A}}(H,H')$ = min$\{\ell_{\mathbf{A}}(i,j): i \in H, j \in H'\}$. Let $\mathbf{C}_H$ = $(C_i : i \in H)$, $\mathcal{L}_d$ be the set of bounded, $\mathbb{R}$-valued, Lipschitz functions on $\mathbb{R}^d$, and 
\begin{equation}
\begin{split}
\nonumber
\mathcal{P}(h,h';s)=\{(H,H'): H, H' \subseteq N_n, |H|=h,|H'|=h', \ell_{\mathbf{A}}(H,H')\ge s \}.
\end{split}
\end{equation}
A triangular array $\{C_i\}_{i=1}^n$ is conditionally $\Psi$-dependence given $\mathcal{F}_n$ if there exist (i) an $\mathcal{F}_n$-measurable sequence $\{ \widetilde{\theta}_{n,s}\}_{s,n \in \mathbb{N}}$ with $\theta_{n,0}$ = 1 $\forall$ $n$ such that $\sup_{n}$ $\widetilde{\theta}_{n,s}$ $\to$ 0 as $s$ $\to$ $\infty$, and (ii)  functionals $\{ \Psi_{h,h'} (\cdot ,\cdot)\}_{h,h'\in \mathbb{N}}$ with $\Psi_{h,h'}$: $\mathcal{L}_h$ $\times$ $\mathcal{L}_h'$ $\to$ $[0, \infty)$ such that
\begin{equation}
\begin{split}
|Cov(f(\mathbf{C}_H),f'(\mathbf{C}_{H'})\mid \mathcal{F}_n)| \le \Psi_{h,h'} (f ,f')\widetilde{\theta}_{n,s}
\end{split}
\end{equation}
for all $n$, $h$, $h'$ $\in$ $\mathbb{N}$; $s$ > 0; $f$ $\in$ $\mathcal{L}_h$; $f'$ $\in$ $\mathcal{L}_{h'}$; and (H, H') $\in$ $\mathcal{P}(h,h';s)$.
\end{definition} 
\par
This work extends temporal $\Psi$-dependence \citep{doukhan1999new} to network settings by replacing temporal distance with shortest-path distance. The $\Psi$-network dependence requires that the covariance between any two observation sets $\mathbf{C}_H$ and $\mathbf{C}_{H'}$ decays to zero as the distance between them increases.
\par
Define 
\begin{equation}
\begin{split}
\nonumber
C_i=\left(\frac{\mathbf{1}_i(g,1)}
{E(\mathbf{1}_i(g,1))}
-
\frac{q^*_1(Z_{i}, \mathbf{X}, \mathbf{A})\mathbf{1}_i(g,0)}
{E(\mathbf{1}_i(g,1))}\right)
\left(\Delta Y_{i}-h^*_1(W_{i}, \mathbf{X}, \mathbf{A})\right)
\end{split}
\end{equation}
and define $\sigma_{n}^{ADT}$ = Var$\left[n^{-1/2}\sum_{i \in N_n}\left(C_i-\tau^{dr}_{ADT}(g)\right) \right]$.
Let $K$ be the constant in Assumption 3.4, $\lfloor s \rfloor$  be $s$ rounded down to the nearest integer, $i^{*}(\cdot)$ be the identity function $x \mapsto x$ on $\mathbb{R}$, and Lip($f$) be the Lipschitz constant and $\|f\|_{\infty}$ = $\sup |f|$.
\begin{theorem}\label{theorem3}
Under Assumptions 3.2-3.5, $\{C_i\}_{i=1}^n$ is conditionally $\Psi$-dependence given $(\mathbf{Z}, \mathbf{X}, \mathbf{A})$ in that (6) holds with $\widetilde{\theta}^{ADT}_{n,s}$ = $\theta^{ADT}_{n,\lfloor s/2 \rfloor}\mathbf{1}\{s > 2max\{K,1\}\}+\mathbf{1}\{s \le 2max\{K,1\}\}$ for all $n$ $\in$ $\mathbb{N}$ and $s$ > 0 and 
\begin{equation}
\begin{split}
\nonumber
 \Psi_{h,h'} (f ,f')=2(||f||_{\infty}||f'||_{\infty}+h||f'||_{\infty}Lip(f)+h'||f||_{\infty}Lip(f'))
\end{split}
\end{equation} 
for  either h, h' $\in$ $\mathbb{N}$, f $\in$ $\mathcal{L}_h$, f' $\in$ $\mathcal{L}_{h'}$, or h = h' =1 and f = f' = $i^*$. 
\end{theorem}
The proof of Theorem~\ref{theorem3} can be found in Appendix. Theorem~\ref{theorem3} shows that $\{C_i\}_{i=1}^n$ is conditionally $\Psi$-dependent, and we can  apply results due to \cite{kojevnikov2021limit} to show that $\hat{\tau}^{mr}_{ADT}(g)$ is  asymptotically normal. 
\par
Let $N^{\partial}_{\mathbf{A}}(i,s)$ = $\{j \in N_n: \ell_{\mathbf{A}}(i,j) =s\}$ be the subset of $N_n$ that are exactly at distance $s$ from unit $i$ $\in$ $N_n$. Then, define $M^{\partial}_{N_n}(s,k)$ = $n^{-1}\sum_{i\in N_n}|N^{\partial}_{\mathbf{A}}(i,s)|^k$, which measures the denseness of $\mathbf{A}$ restricted on $N_n$. When $k$ = 1, we denote $M^{\partial}_{N_n}(s)$ = $M^{\partial}_{N_n}(s,1)$. Furthermore,  let $\Delta_{N_n}(s, m; k)$ = $n^{-1}\sum_{i \in N_n}max_{j \in N^{\partial}_{\mathbf{A}}(i,s)}|N_{\mathbf{A}}(i,m)\backslash N_{\mathbf{A}}(j,s-1)|^k$, where $N_{\mathbf{A}}(j,s-1)$ = $\varnothing$ if $s$ = 0. This represents the $k$th sample moment of the maximum number (across all $j$ at distance s from $i$) of units who are within distance $m$ from $i$ but at least distance $s$ apart from $j$. In addition, we define $c_{N_n}(s,m;k)$ = $inf_{\alpha > 1}$$[\Delta_{N_n}(s, m; k\alpha)]^{\frac{1}{\alpha}}$$[M^{\partial}_{N_n}(s,\alpha/(\alpha-1))]^{1-\frac{1}{\alpha}}$. This quantity measures the denseness of the network, which plays an important role in establishing the CLT \citep{hoshino2024causal}. 
 \begin{assumption}[Weak dependence for ADT]. (a) $max_{1\le s \le 2K}M^{\partial}_{N_n}(s)= O(1)$, where K is as given in Assumption 3.4. (b)  There exist some positive sequence $m_n$ $\to$ $\infty$ and a constant 0 < $\varepsilon$ < 1 such that for each $k$ $\in$ $\{1,2\}$, $n^{-k/2}$$(\sigma_{N_n}^{ADT})^{-(2+k)}\sum_{s=0}^{n-1}$$c_{N_n}(s,m_n;k)$$(\widetilde{\theta}^{ADT}_{n,s})^{1-\varepsilon}$ $\to$ 0, $n^{k/2}(\sigma_{N_n}^{ADT})^{-k}(\widetilde{\theta}^{ADT}_{n,m_n})^{1-\varepsilon}$ $\to$ 0, and ${lim\, sup}_{n \to \infty}\sum_{s=0}^{n-1}M^{\partial}_{N_n}(s,2)^{1/2}(\widetilde{\theta}^{ADT}_{n,s})^{1-\varepsilon}$ < $\infty$ a.s. 
 \end{assumption}
Assumption 3.6(a) rules out that there are a non-negligible proportion of units whose $2K$ neighborhoods may grow to infinity as $n$ increases. If one assumes that each individual can hold only a limited number of interacting partners, Assumption 3.6(a) is satisfied with $M_{S_n}^\partial (s, k) < \infty$ for all $s, k < \infty$. However, for example, it is violated if the network is a complete graph. The first two terms in Assumption 3.6(b) correspond to Assumption 3.4 of \cite{kojevnikov2021limit}, which they utilize to establish a CLT. The third is similar with \cite{leunggnn} and used to asymptotically linearize our robust estimator under network
dependence. 
\begin{theorem}\label{theorem4}
Under Assumptions 3.1-3.6, 
\begin{equation}
\begin{split}
\nonumber
(\sigma_n^{ADT})^{-1/2}\sqrt{n}(\hat{\tau}^{dr}_{ADT}(g)-\tau^{dr}_{ADT}(g))\overset{d}{\rightarrow} \mathcal{N}(0, 1).
\end{split}
\end{equation} 
\end{theorem}
The proof of Theorem~\ref{theorem4} can be find in Appendix A. 
\par
The next result characterizes the asymptotic properties of $\hat{\sigma}_n^{ADT}$. We consider inference methods based on network HAC estimation and demonstrate that the HAC estimator exhibits asymptotic biases. This bias arises from the inability to estimate the heterogeneous means that appear in the asymptotic variances in Theorem~\ref{theorem4}. This is a well-known issue in the design-based uncertainty framework \citep{imbens2015causal}.
Define
\begin{equation}
\begin{split}
\nonumber
\mathcal{J}_n(s,m)=\{(i,j,k,l)\in N_n^4: k\in N(i,m),l \in (j,m),\ell_{\mathbf{A}}(i,j) =s \}.
\end{split}
\end{equation} 
 \begin{assumption}[HAC]. (a)  For some constant $C$ > 0 and all $i$ $\in$ $N_n$,  let $|max(\hat{h}^*_1( W_{i}, \mathbf{X}, \mathbf{A}), \hat{q}^*_1( Z_{i}, \mathbf{X}, \mathbf{A}), h^*_1( W_{i}, \mathbf{X}, \mathbf{A}), q^*_1( Z_{i}, \mathbf{X}, \mathbf{A}))|$ < $C$ $a.s.$, $1/n\sum_{i\in N_n}(\hat{h}^*_1( W_{i}, \mathbf{X}, \mathbf{A})-h^*_1( W_{i}, \mathbf{X}, \mathbf{A}))^2$ = $o_P(n^{-1/2})$ and $1/n\sum_{i\in N_n}(\hat{q}^*_1( Z_{i}, \mathbf{X}, \mathbf{A})-q^*_1( Z_{i}, \mathbf{X}, \mathbf{A}))^2$ = $o_P(n^{-1/2})$. (b) For some constant $\varepsilon$ $\in$ (0,1) and $b_n$ $\to$ $\infty$, $\text{lim}_{n \to \infty}n^{-1}\sum_{s=0}^{\infty}c_n(s,b_n;2)(\widetilde{\theta}^{ADT}_{n,s})^{1-\varepsilon}$ = 0 $a.s.$ (c) $n^{-1}\sum_{i=1}^n|N_{\mathbf{A}}(i,b_n)|$ = $o_P(\sqrt{n})$. (d) $n^{-1}\sum_{i=1}^n|N_{\mathbf{A}}(i,b_n)|^2$ = $O_P(\sqrt{n})$. (e) $\sum_{s=0}^n|\mathcal{J}_n(s,b_n)|\widetilde{\theta}^{ADT}_{n,s}$ = $o(n^2)$. 
 \end{assumption}
Part (a) provides a mild strengthening of Assumption 3.1, as all nuisance functions are estimated nonparametrically. Since it does not require uniform convergence, it is easier to verify for machine learning estimators. Parts (b)-(e) restrict both the network structure and the rate of divergence of $b_n$ in a similar manner to Assumption 7 of \cite{leunggnn} and Assumption 4.1 of \cite{kojevnikov2021limit}. 
\par
Define
\begin{equation}
\begin{split}
\nonumber
\hat{\sigma}_n^{ADT}
 = \frac{1}{n}\sum_{i \in N_n}\sum_{j \in N_n}
\hat{\tau}_{ADT,i}(g)\hat{\tau}_{ADT,j}(g) \mathbf{1}\{\ell_{\mathbf{A}}(i,j) \le b_n\},
\end{split}
\end{equation} 
where
\begin{equation}
\begin{split}
\nonumber
\hat{\tau}_{ADT,i}(g)=
\left(\frac{\mathbf{1}_i(g,1)}
{\hat{E}(\mathbf{1}_i(g,1))}
-
\frac{\hat{q}^*_1(Z_{i}, \mathbf{X}, \mathbf{A})\mathbf{1}_i(g,0)}
{\hat{E}(\mathbf{1}_i(g,1))}\right)
\left(\Delta Y_{i}-\hat{h}^*_1(W_{i}, \mathbf{X}, \mathbf{A})\right)
-\hat{\tau}^{dr}_{ADT}(g).
\end{split}
\end{equation}
Similarly, define
\begin{equation}
\begin{split}
\nonumber
\hat{\sigma}_n^{ADT*}
 = \frac{1}{n}\sum_{i \in N_n}\sum_{j \in N_n}
\tilde{\tau}_{ADT,i}(g)\tilde{\tau}_{ADT,j}(g) \mathbf{1}\{\ell_{\mathbf{A}}(i,j) \le b_n\},
\end{split}
\end{equation} 
where
$\tilde{\tau}_{ADT,i}(g)$ = $C_i -\mathrm{E}[C_i]$, and
 \begin{equation}
\begin{split}
\nonumber
B_n
 =& \frac{1}{n}\sum_{i \in N_n}\sum_{j \in N_n}
(C_i -\tau^{dr}_{ADT}(g))(C_j -\tau^{dr}_{ADT}(g))\mathbf{1}\{\ell_{\mathbf{A}}(i,j) \le b_n\}.
\end{split}
\end{equation} 
\begin{theorem}\label{th4}
Under Assumption 3.7 and the assumptions of Theorem 3,
 \begin{equation}
\begin{split}
\nonumber
\hat{\sigma}_n^{ADT} = \hat{\sigma}_n^{ADT*}+B_n+o_P(1)
\quad \text{and}\quad |\hat{\sigma}_n^{ADT*}-\sigma_n^{ADT}|
\overset{p}{\rightarrow} 0.
\end{split}
\end{equation} 
\end{theorem}
The proof of Theorem~\ref{th4} can be find in Appendix A.6. In this article, we apply the bandwidth
 \begin{equation}
\begin{split}
b_n=\lceil \tilde b_n \rceil\quad\text{for}\quad
\tilde b_n=
\begin{cases}
\dfrac{1}{4}\,\mathcal{L}(\mathbf{A}), & \text{if }\mathcal{L}(\mathbf{A})<2^{\frac{\log n}{\log\delta(\mathbf{A})}},\\[4pt]
\mathcal{L}(\mathbf{A})^{1/4}, & \text{otherwise,}
\end{cases}
\end{split}
\end{equation} 
where $\lceil  \cdot \rceil$ rounds up to the nearest integer, $\delta(\mathbf{A})$ = $n^{-1}\sum_{i,j} A_{ij}$ is the average degree,
and $\mathcal{L}(\mathbf{A})$ = $(n(n-1))^{-1}\sum_{i\ne j} \ell_{\mathbf{A}}(i,j)$ is the average path length, which means the average over all unit pairs in the largest component of $\mathbf{A}$. A component is a connected subnetwork such that
all units in the subnetwork have infinite path distance to non-members of the subnetwork. This is identical to the bandwidth selection in \cite{leunggnn}, which verifies the high-level assumptions required to characterize the asymptotic properties of $\hat{\sigma}_n^{ADT} $ under the bandwidth choice (7). 

\section{Simulation study}\label{sec4}
In this section, we conduct Monte Carlo simulations to evaluate the finite-sample properties of our proposed estimator. Its performance is compared against conventional regression models designed to account for spillover effects. The data generation process (DGP) is specified as follows: First, an undirected network $\mathbf{A}$ is generated using the Erd\H{o}s--R\'enyi model, setting the connection probability between any two distinct nodes to $0.5$. Simulations are run for network sizes $n \in \{1500, 2000, 2500, 3000\}$. For each node $i$, 15 covariates are generated from a multivariate normal distribution, $\mathcal{N}(\mathbf{0},\mathbf{I_{15\times 15}})$, and partitioned into 10-dimensional observed covariates $\mathbf{X_i}$ and 5-dimensional unobserved confounders $\mathbf{U_i}$. Finally, independent error terms $(e_i, \eta_i, \vartheta_i)$ are drawn from $\mathcal{N}(0, 0.1)$.
\par
A critical feature of our DGP is the incorporation of network structures into the generation of key variables. The treatment assignment $(D_i)_{i=1}^n$, the outcomes $(Y_{i0}, Y_{i1})_{i=1}^n$, and double negative controls variables $(W_{i}, Z_{i})_{i=1}^n$ are all functions of not only the ego's covariates ($\mathbf{X_i}, \mathbf{U_i}$)  but also their neighbors' covariates. Let $\beta_X$ = $(0.1, 0.5, 0.5^2, 0.5^3, 0.5^4, 0, 0, 0, 0, 0)$ and $\beta_U$ = $(0.1, 0.5, 0.5^2, 0, 0)$, and define
 \begin{equation}
\begin{split}
\nonumber
Z_i=0.1+0.7\frac{\sum_{j=1}^{n-1}A_{ij}U_{j1}}{\sum_{j=1}^{n-1}A_{ij}}
+0.3\frac{\sum_{j=1}^{n-1}A_{ij}X_{j1}}{\sum_{j=1}^{n-1}A_{ij}}
+0.2\beta_X \mathbf{X_i}+0.3\beta_U \mathbf{U_i} +\eta_i,
\end{split}
\end{equation} 
 \begin{equation}
\begin{split}
\nonumber
W_i=0.1+0.9\frac{\sum_{j=1}^{n-1}A_{ij}U_{j1}}{\sum_{j=1}^{n-1}A_{ij}}
+0.7\frac{\sum_{j=1}^{n-1}A_{ij}X_{j1}}{\sum_{j=1}^{n-1}A_{ij}}
+0.3\beta_X \mathbf{X_i}+0.4\beta_U \mathbf{U_i} +\vartheta_i,
\end{split}
\end{equation} 
 \begin{equation}
\begin{split}
\nonumber
D_i = \mathbf{1}\{0.1+0.8\frac{\sum_{j=1}^{n-1}A_{ij}U_{j1}}{\sum_{j=1}^{n-1}A_{ij}}
+0.5\frac{\sum_{j=1}^{n-1}A_{ij}X_{j1}}{\sum_{j=1}^{n-1}A_{ij}}
+0.4\beta_X \mathbf{X_i}+0.5\beta_U \mathbf{U_i} 
+\frac{\sum_{j=1}^{n-1}A_{ij}e_j}{\sum_{j=1}^{n-1}A_{ij}}+e_i
> 0
\},
\end{split}
\end{equation} 
 \begin{equation}
\begin{split}
\nonumber
Y_{i0}=0.1+0.6\frac{\sum_{j=1}^{n-1}A_{ij}U_{j1}}{\sum_{j=1}^{n-1}A_{ij}}
+0.4\frac{\sum_{j=1}^{n-1}A_{ij}X_{j1}}{\sum_{j=1}^{n-1}A_{ij}}
+0.5\beta_X \mathbf{X_i}+0.6\beta_U \mathbf{U_i}+\frac{\sum_{j=1}^{n-1}A_{ij}e_j}{\sum_{j=1}^{n-1}A_{ij}}+e_i,
\end{split}
\end{equation} 
 \begin{equation}
\begin{split}
\nonumber
Y_{i1}=Y_{i0}+\tau D_i + 0.2(\frac{\sum_{j=1}^{n-1}A_{ij}U_{j1}}{\sum_{j=1}^{n-1}A_{ij}})^2
&+ 0.2(\frac{\sum_{j=1}^{n-1}A_{ij}X_{j1}}{\sum_{j=1}^{n-1}A_{ij}})^2
\\&+0.5 \frac{\sum_{j=1}^{n-1}A_{ij}D_{j}}{\sum_{j=1}^{n-1}A_{ij}}
+0.2 \beta_X \mathbf{X_i}+0.3 \beta_U \mathbf{U_i}+0.3W_i,
\end{split}
\end{equation}
and $\Delta Y_i$ = $Y_{i1}$ - $Y_{i0}$. 
The above models imply that the ADT is $\tau$, which we set to be $0.5$. 
\par
Next, the basic settings for the GNN estimation are explained. Specifically, an exposure mapping is first intentionally misspecified as $G_i=\mathbf{1}\{\sum_{j=1}^{n-1}A_{ij}D_j>0\}$ (the correct specification being $G_i=\sum_{j=1}^{n-1}A_{ij}D_j/\sum_{j=1}^{n-1}A_{ij}$), and $\tau(1)$ is then estimated. Our GNNs model utilizes the principal neighborhood aggregation (PNA) architecture from \cite{corso2020principal}. This architecture enhances expressive power by integrating multiple aggregators, making it more robust in networks where node degrees vary widely. In particular, we employ ``mean'' and ``sum'' aggregators to combine information from neighboring nodes. To stabilize the aggregation process with respect to node degree, these aggregators are paired with three degree scalers: ``identity'', ``amplification'', and ``attenuation''. The ``amplification'' and ``attenuation'' scalers are based on the logarithm of the node degree, a design that helps prevent the exponential amplification of gradients across successive GNNs layers. 
\par
All GNNs estimators are set with $L = 1$. The update function within each layer consists of a MLP with a hidden layer dimension of 16, using $\mathbf{ReLU}$ as the activation function. For model training, we employ the Adam variant of stochastic gradient descent for optimization. The learning rate is fixed at 0.01, and the number of training epochs for each nuisance function is 500. All neural network implementations are based on the $\mathbf{torch}$ package in $\mathbf{R}$.
\par
Subsequently, we compare the proposed estimator with the following model:
 \begin{equation}
\begin{split}
\Delta Y_i = \alpha + \tau D_i + \delta \frac{\sum_{j=1}^{n-1}A_{ij}D_j}{\sum_{j=1}^{n-1}A_{ij}} + \beta^{\text{T}} \mathbf{X_i}
+\gamma^{\text{T}} \frac{\sum_{j=1}^{n-1}A_{ij}\mathbf{X_j}}{\sum_{j=1}^{n-1}A_{ij}}+\varepsilon_i, 
\end{split}
\end{equation}
where both $\beta$ and $\gamma$  are 10-dimensional column vectors. This model uses the correct exposure mapping $G_i=\sum_{j=1}^{n-1}A_{ij}D_j/\sum_{j=1}^{n-1}A_{ij}$ and accounts for the covariates of neighbors. 
\par
Table~\ref{tab1} compares our proposed estimator ($\tau_{ADT}(1)$) and Model (9) for sample sizes $n = 1500$ to $3000$. Despite a deliberately misspecified exposure mapping, our estimator demonstrates consistency: as $n$ increases, both bias and RMSE decrease (RMSE drops from 0.3240 to 0.2921), indicating robustness to unmeasured confounding. In contrast, Model (9), even with the correct exposure mapping, exhibits persistent bias ($\approx 0.37$) and stable RMSE ($\approx 0.373$) across all sample sizes. This lack of convergence confirms its inability to address omitted variable bias from $\mathbf{U_i}$, yielding precise but biased estimates.
\begin{table}
\centering
\begin{threeparttable}
\caption{Monte Carlo Simulation Results}
\label{tab1}
\begin{tabular}{lcccccccc} 
\toprule
           & \multicolumn{4}{c}{$\tau_{ADT}(1)$} & \multicolumn{4}{c}{Model (9)} \\
\cmidrule(r){2-5} \cmidrule(l){6-9} 
n          & 1500    & 2000    & 2500   & 3000   & 1500   & 2000   & 2500   & 3000   \\
\midrule
Estimate   & 0.7156  & 0.7056  & 0.6801 & 0.6924 & 0.8733 & 0.8730 & 0.8732 & 0.8735 \\
SE         & 0.2423  & 0.2247  & 0.2105 & 0.2200 & 0.0094 & 0.0082 & 0.0078 & 0.0067 \\
Bias       & 0.2156  & 0.2056  & 0.1801 & 0.1924 & 0.3733 & 0.3730 & 0.3732 & 0.3735 \\
RMSE       & 0.3240  & 0.3044  & 0.2768 & 0.2921 & 0.3734 & 0.3731 & 0.3733 & 0.3735 \\
\bottomrule
\end{tabular}
\begin{tablenotes}
  \small 
  \item[*] \textit{Notes:} The true value of the parameter is $\tau = 0.5$. Results are based on 500 Monte Carlo simulations.

\end{tablenotes}
\end{threeparttable}
\end{table}

\par
In summary, the simulation results in Table~\ref{tab1} provide evidence for the superiority of our proposed method. Even when subject to a misspecified exposure mapping, our GNN-based doubly robust estimator yields significantly lower bias and RMSE compared to the conventional regression-based approach.
\section{Empirical application} \label{sec5}
In this section, we apply our methodology to estimate the causal effect of China's 2012 Green Credit Policy on corporate green innovation \citep{huang2023green,li2024impact}. Crucially, this evaluation necessitates explicitly accounting for \textbf{unmeasured network confounding}, as corporate responses are often driven by latent factors?such as informal political connections and executive networks?that influence both regulatory treatment and innovation. Relying solely on observables would likely yield biased estimates. Moreover, given the complex, non-linear nature of knowledge spillovers, our framework integrating double negative controls with GNNs is uniquely suited to robustly proxy for these latent confounders and flexibly model high-dimensional network interactions.
\par
We analyze data from Chinese A-share listed companies in 2011 (pre-policy) and 2013 (post-policy), sourced from the China National Intellectual Property Administration (CNIPA) and the China Stock Market $\&$ Accounting Research (CSMAR) Database. After excluding financial institutions and firms with data irregularities (e.g., ST stocks), the final sample comprises 1,664 observations. The network adjacency matrix is defined as $A_{ij} = 1$ if firms share the same industry or are located in adjacent provinces. The outcome $Y_{it}$ is measured as the natural logarithm of granted green patents plus one. The treatment variable $D_{i}$ is defined as follows:
\begin{align}
D_{i} &= \begin{cases}
    1, & \text{if enterprise } i \text{ belongs to the six major heavily polluting industries}, \\
    0, & \text{if enterprise } i \text{ does not belong to the six major heavily polluting industries},
\end{cases} \nonumber
\end{align}
where the six major heavily polluting industries include thermal power, iron and steel, petrochemicals, cement, non-ferrous metals, and chemicals \citep{liu2019green}. Based on the treatment status $\mathbf{D}$, let $W$ (the NCO) and $Z$ (the NCE) be the proportion of treated individuals among an individual's first-order neighbors and non-neighbors, respectively. Notably, this choice of a double-negative control is consistent with Assumption 2.3. In addition, we include 10 control variables in the model, including the proportion of institutional investor shareholding (InstHold), duality of CEO and chairman (Duality), proportion of independent directors (IndDir), ratio of owner's equity to market value (EqToMV), asset-liability ratio (Lev), ratio of cash to total assets (CashTA), ratio of total fixed assets to total assets (FixTA), return on total assets (ROA), number of employees (EmpNum), and Tobin's Q (TobinQ). Table~\ref{tab2} reports the descriptive statistics of the main variables used in the study.
\begin{table}[h!]
  \centering
  \caption{Descriptive Statistics}
  \begin{tabular}{lrrrrrrr}
    \toprule
    Variables & Mean    & St.dev  & Min     & p25     & p50     & p75     & Max      \\
    \midrule
    $\Delta$Y & 0.0409  & 0.5843  & -2.9444 & 0.0000  & 0.0000  & 0.0000  & 3.2581   \\
    D         & 0.0270  & 0.1623  & 0.0000  & 0.0000  & 0.0000  & 0.0000  & 1.0000   \\
    W         & 0.0242  & 0.0262  & 0.0000  & 0.0096  & 0.0124  & 0.0325  & 0.1833   \\
    Z         & 0.0301  & 0.0042  & 0.0205  & 0.0258  & 0.0324  & 0.0343  & 0.0371   \\
    InstHold  & 0.4849  & 0.2367  & 0.0001  & 0.3134  & 0.5063  & 0.6672  & 0.9826   \\
    Duality   & 0.2242  & 0.4172  & 0.0000  & 0.0000  & 0.0000  & 0.0000  & 1.0000   \\
    IndDir    & 0.3701  & 0.0590  & 0.1667  & 0.3333  & 0.3333  & 0.4000  & 0.7143   \\
    EqToMV    & 0.3324  & 0.1712  & -0.0246 & 0.2016  & 0.3000  & 0.4427  & 1.0000   \\
    Lev       & 0.2018  & 0.2212  & 0.0018  & 0.0872  & 0.1524  & 0.2712  & 0.9239   \\
    CashTA    & 0.4400  & 0.1593  & 0.0071  & 0.2632  & 0.4512  & 0.6104  & 1.2517   \\
    FixTA     & 0.2248  & 0.1662  & 0.0002  & 0.0968  & 0.1932  & 0.3212  & 0.9709   \\
    ROA       & 0.0473  & 0.0526  & -0.4041 & 0.0212  & 0.0426  & 0.0712  & 0.3020   \\
    EmpNum    & 4269.7  & 17280.7 & 13      & 871.5   & 1890    & 4151.8  & 552810   \\
    TobinQ    & 1.689   & 0.8414  & 0.759   & 1.198   & 1.433   & 1.862   & 9.548    \\
    \bottomrule
  \end{tabular}
  \label{tab2}%
\end{table}
\par
Next, we select different exposure mappings to explore the impact of GCP on enterprises' green innovation in Table~\ref{tab3}. The first mapping $G_{1i}=\mathbf{1}\{\sum_{j=1}^{n-1}A_{ij}D_j>0\} = 1$ indicates that there exists a treated group around the enterprise. The second mapping $G_{2i}=\mathbf{1}\{\sum_{j=1}^{n-1}A_{ij}D_j>1\} = 1$ indicates that there are at least two treated groups around the enterprise. And the third mapping  $G_{3i} = \mathbf{1} \left( \sum_{j=1}^{N} A_{ij} D_j > \frac{1}{5} \left[ \frac{1}{N} \sum_{k=1}^{N} \left( \sum_{j=1}^{N} A_{kj} D_j \right) \right] \right)=1$ indicates that the number of treated neighbors of an enterprise is greater than one-fifth of the average number of treated neighbors across all enterprises. The first column, $ADT(G_{1i}=1)$, estimates the direct policy effect for the subset of treated firms that have at least one treated neighbor. The resulting ADT is 0.2662 and is highly statistically significant (P-value = 0.0014). This indicates that the GCP had a significant direct impact on heavily polluting firms, strongly incentivizing or compelling them to pursue green innovation, particularly when they were networked with at least one other polluting firm. The second column, $ADT(G_{2i}=1)$, focuses on treated firms with at least two treated neighbors. The estimated ADT increases in magnitude to 0.3074. Numerically, this suggests that the direct effect of the policy is even stronger for treated firms embedded in denser clusters of their peers. Finally, the third column, $ADT(G_{3i}=1)$, uses a relative exposure measure. It shows the ADT for treated firms whose number of treated neighbors is greater than one-fifth of the network average. The estimate is 0.1405 and remains statistically significant at the 5$\%$ level (P-value = 0.0251). This confirms that the significant direct effect of the GCP on polluting firms is robust, particularly for those located in relatively ``dense'' clusters of other treated enterprises.

\begin{table}[htbp] 
  \centering
  \caption{ Estimated results}
  \label{tab3} 
  \begin{tabular}{lccc}
    \toprule
    & $ADT(G_{1i}=1)$ & $ADT(G_{2i}=1)$ & $ADT(G_{3i}=1)$ \\
    \midrule
    Estimate    & 0.2662 & 0.3074 & 0.1405 \\
    SE          & 0.0836 & 0.1834 & 0.0627 \\
    Sample size & 1637   & 1637   & 1575   \\
    $b_n$       & 2      & 2      & 2      \\
    P-value     & 0.0014 & 0.0938 & 0.0251 \\
    \bottomrule
  \end{tabular}
\end{table}
\par
In summary, the results in Table~\ref{tab3} reveal significant heterogeneity in the direct effect of the GCP. The policy's direct impact on promoting green innovation among polluting firms is robust and significant, and this direct effect appears to be (at least in magnitude) stronger for treated firms that are more deeply exposed to other treated firms within their network.
\section{Conclusion}\label{sec6}
Estimating causal effects from observational network data is challenging due to unmeasured network confounding. Traditional methods, including standard DID, often fail in these settings. Therefore, we propose a general DID framework integrating DNC and GNNs. We establish an identification strategy founded upon the latent parallel trends assumption, leveraging DNC via confounding bridge functions. This approach yields doubly robust, non-parametric identification of the treatment effects, requiring only one of the two bridge functions to be correctly specified for identification to hold. 
Subsequently, we develop a doubly robust estimator. This estimator leverages GNNs combined with the GMM to flexibly estimate the high-dimensional bridge functions inherent in the DNC approach. We further establish the estimator's asymptotic normality under conditions of $\psi$-network dependence and ANI, and provide a network HAC variance estimator. Simulation studies confirm the favorable finite-sample performance of our estimator relative to conventional models, even under misspecification of the exposure mapping. Finally, an empirical application evaluating the impact of China's green credit policy demonstrate the method's practical utility.

\section*{Conflict of interest}  
On behalf of all authors, the corresponding author states that there is no conflict of interest.

\section*{Funding} 
This work was supported by the [National Social Science Fund of China], [24BTJ064].

\begin{center}

{\large\bf SUPPLEMENTARY MATERIAL}

\end{center}

\begin{description}
\item[Title:]
Difference-in-Differences using Double Negative Controls and Graph Neural
Networks for Unmeasured Network Confounding
\end{description}
\section*{A Proofs}
\subsection*{A.1 Proof of Theorem 1}
This proof adopts the methodology of \cite{miao2018identifying} and adapts it to our DID framework under network interference. First, we will prove
\begin{equation}
\begin{split}
\nonumber
&\quad E[Y_{i1}(1,\mathbf{D_{-i}})-Y_{i1}(0,\mathbf{D_{-i}})
|G_{i}=g, D_{i}=1]=
E[\Delta Y_i
-h_1(W_{i}, \mathbf{X}, \mathbf{A})
|G_{i}=g, D_{i}=1].
\end{split}
\end{equation}
Under Assumption 2.1 and 2.2, 
\begin{equation}
\begin{split}
\nonumber
& E\big[Y_{i1}(1,\mathbf{D}_{-i}) - Y_{i1}(0,\mathbf{D}_{-i}) 
   \,\big|\, G_{i}=g, D_{i}=1\big] \\
&= E\Big[E\big[Y_{i1}(1,\mathbf{D}_{-i}) - Y_{i0}(0,\mathbf{D}_{-i}) 
   \,\big|\, G_{i}=g, D_{i}=1, \mathbf{U}, \mathbf{X}, \mathbf{A}\big] \\
&\quad - E\big[Y_{i1}(0,\mathbf{D}_{-i}) - Y_{i0}(0,\mathbf{D}_{-i}) 
   \,\big|\, G_{i}=g, D_{i}=0, \mathbf{U}, \mathbf{X}, \mathbf{A}\big] 
   \,\Big|\, G_{i}=g, D_{i}=1\Big] \\
&= E\Big[E\big[\Delta Y_i 
   \,\big|\, G_{i}=g, D_{i}=1, \mathbf{U}, \mathbf{X}, \mathbf{A}\big] \\
&\quad - E\big[\Delta Y_i 
   \,\big|\, G_{i}=g,D_{i}=0,  \mathbf{U}, \mathbf{X}, \mathbf{A}\big] 
   \,\Big|\, G_{i}=g, D_{i}=1\Big],
\end{split}
\end{equation}
where the first equality follows from Assumption 2.1 and the second equality follows from Assumption 2.2. Under Assumption 2.3,
\begin{equation}
\begin{split}
\nonumber
&  E\big[\Delta Y_i
-h_1(W_{i}, \mathbf{X}, \mathbf{A}) 
   \,\big|\, G_{i}=g, D_{i}=1\big] \\
&= E\Big[E\big[\Delta Y_i 
   \,\big|\, G_{i}=g, D_{i}=1, \mathbf{U}, \mathbf{X}, \mathbf{A}\big] \\
&\quad - E\big[h_1(W_{i},  \mathbf{X}, \mathbf{A})
   \,\big|\, G_{i}=g, D_{i}=0, \mathbf{U}, \mathbf{X}, \mathbf{A}\big] 
   \,\Big|\, G_{i}=g, D_{i}=1\Big] 
, 
\end{split}
\end{equation}
where the equality follows from Assumption 2.3(a). 
\par
Under Assumption 2.4, $E[\Delta Y_{i}|G_{i}=g, D_{i}=d, \mathbf{U}, \mathbf{X}, \mathbf{A}]
=
E[h_1( W_{i}, \mathbf{X}, \mathbf{A})|G_{i}=g,D_{i}=0,  \mathbf{U}, \mathbf{X},\mathbf{A}]$, we get
\begin{equation}
\begin{split}
\nonumber
&\quad E[Y_{i1}(1,\mathbf{D_{-i}})-Y_{i1}(0,\mathbf{D_{-i}})
|G_{i}=g, D_{i}=1]=
E[\Delta Y_{i}
-h_1(W_{i}, \mathbf{X}, \mathbf{A})
|G_{i}=g, D_{i}=1].
\end{split}
\end{equation}
Next, we prove that the confounding bridge function is identified by
\begin{equation}
\begin{split}
\nonumber
E[\Delta Y_{i}|G_{i}=g, D_{i}=0, Z_{i}, \mathbf{X}, \mathbf{A}]
=
E[h_1(W_{i}, \mathbf{X}, \mathbf{A})|G_{i}=g, D_{i}=0, Z_{i}, \mathbf{X}, \mathbf{A}].  
\end{split}
\tag{A.1}
\end{equation}
using Assumption 2.3, we have
\begin{equation}
\begin{split}
\nonumber
&\quad E\left[\Delta Y_{i}|G_{i}=g, D_{i}=0, Z_{i}, \mathbf{X}, \mathbf{A}\right]\\
&=E\left[E\left[\Delta Y_{i}|G_{i}=g, D_{i}=0, Z_{i}, \mathbf{U},\mathbf{X}, \mathbf{A}\right]
|G_{i}=g, D_{i}=0, Z_{i}, \mathbf{X}, \mathbf{A}\right]\\
&=E\left[E\left[\Delta Y_{i}|G_{i}=g, D_{i}=0, \mathbf{U},\mathbf{X}, \mathbf{A}\right]
|G_{i}=g, D_{i}=0, Z_{i}, \mathbf{X}, \mathbf{A}\right].\\
\end{split}
\end{equation}
Similarly, we have 
\begin{equation}
\begin{split}
\nonumber
&\quad E\left[h_1(W_{i},  \mathbf{X}, \mathbf{A})|G_{i}=g, D_{i}=0, Z_{i}, \mathbf{X}, \mathbf{A}\right]\\
&=E\left[E\left[h_1(W_{i},  \mathbf{X}, \mathbf{A})|G_{i}=g, D_{i}=0, Z_{i}, \mathbf{U},\mathbf{X}, \mathbf{A}\right]
|G_{i}=g, D_{i}=0, Z_{i}, \mathbf{X}, \mathbf{A}\right]\\
&=E\left[E\left[h_1(W_{i},  \mathbf{X}, \mathbf{A})|G_{i}=g, D_{i}=0, \mathbf{U},\mathbf{X}, \mathbf{A}\right]
|G_{i}=g, D_{i}=0, Z_{i}, \mathbf{X}, \mathbf{A}\right].\\
\end{split}
\end{equation}
Then, under Assumption 2.4, we proof
\begin{equation}
\begin{split}
\nonumber
E[\Delta Y_{i}|G_{i}=g, D_{i}=0, Z_{i}, \mathbf{X}, \mathbf{A}]
=
E[h_1(W_{i}, \mathbf{X}, \mathbf{A})|G_{i}=g, D_{i}=0, Z_{i}, \mathbf{X}, \mathbf{A}].
\end{split}
\end{equation}
Finally, under Assumption 2.5, we show that the unique solution to equation (A.1) identifies the outcome confounding bridge function $h$
. Suppose there are two functions $h_1(W_{i}, \mathbf{X}, \mathbf{A})$ and $h'_1(W_{i}, \mathbf{X}, \mathbf{A})$ that satisfy equation (A.1). Then,
\begin{equation}
\begin{split}
\nonumber
E[h_1(W_{i}, g, \mathbf{X}, \mathbf{A})
-h'_1(W_{i}, g, \mathbf{X}, \mathbf{A})
|G_{i}=g, D_{i}=0, Z_{i}=z, \mathbf{X}= \mathbf{x}, \mathbf{A}]=0
\end{split}
\end{equation}
for all g, $\mathbf{x}$, $\mathbf{A}$, and almost all $z$.  Then, under Assumption 2.5, $h_1(W_{i}, g, \mathbf{X}, \mathbf{A})$ = $h'_1(W_{i}, g, \mathbf{X}, \mathbf{A})$ almost surely. Thus, the solution to equation (A.1) identifies the outcome confounding bridge function.

\subsection*{A.2 Proof of Theorem 2}
Similar with Theorem 1, we first prove
\begin{equation}
\begin{split}
\nonumber
&\quad E[Y_{i1}(1,\mathbf{D_{-i}})-Y_{i1}(0,\mathbf{D_{-i}})
|G_{i}=g, D_{i}=1]=
E[\frac{\mathbf{1}_i(g,1)\Delta Y_{i}}
{E(\mathbf{1}_i(g,1))}]
-
E[\frac{q_1(Z_{i}, \mathbf{X}, \mathbf{A})\mathbf{1}_i(g,0)\Delta Y_{i}}
{E(\mathbf{1}_i(g,1))}].
\end{split}
\end{equation}
Under Assumption 2.1 and 2.2, 
\begin{equation}
\begin{split}
\nonumber
& \quad E\big[Y_{i1}(1,\mathbf{D}_{-i}) - Y_{i1}(0,\mathbf{D}_{-i}) 
   \,\big|\, G_{i}=g, D_{i}=1\big] \\
&= E\Big[E\big[Y_{i1}(1,\mathbf{D}_{-i}) - Y_{i0}(0,\mathbf{D}_{-i}) 
   \,\big|\, G_{i}=g, D_{i}=1, \mathbf{U}, \mathbf{X}, \mathbf{A}\big] \\
&\quad - E\big[Y_{i1}(0,\mathbf{D}_{-i}) - Y_{i0}(0,\mathbf{D}_{-i}) 
   \,\big|\, G_{i}=g, D_{i}=0, \mathbf{U}, \mathbf{X}, \mathbf{A}\big] 
   \,\Big|\, G_{i}=g, D_{i}=1\Big] \\
&= E\Big[E\big[\Delta Y_i 
   \,\big|\, G_{i}=g, D_{i}=1, \mathbf{U}, \mathbf{X}, \mathbf{A}\big] \\
&\quad - E\big[\Delta Y_i 
   \,\big|\, G_{i}=g,D_{i}=0,  \mathbf{U}, \mathbf{X}, \mathbf{A}\big] 
   \,\Big|\, G_{i}=g, D_{i}=1\Big]\\
&=E[\frac{\mathbf{1}_i(g,1)\Delta Y_{i}}
{E(\mathbf{1}_i(g,1))}]
-E[\frac{E[\mathbf{1}_i(g,1)|\mathbf{U}, \mathbf{X}, \mathbf{A}]}
{E[\mathbf{1}_i(g,0)|  \mathbf{U}, \mathbf{X}, \mathbf{A}]}
\frac{\mathbf{1}_i(g,0)\Delta Y_i }{E(\mathbf{1}_i(g,1))}
],
\end{split}
\end{equation}
where the first equality follows from Assumption 2.1, the second equality follows from Assumption 2.2, and the last  follows from the law of iterated expectations. Under Assumption 2.3,
\begin{equation}
\begin{split}
\nonumber
&\quad E[\frac{\mathbf{1}_i(g,1)\Delta Y_{i}}
{E(\mathbf{1}_i(g,1))}]
-
E[\frac{q_1(Z_{i}, \mathbf{X}, \mathbf{A})\mathbf{1}_i(g,0)\Delta Y_{i}}
{E(\mathbf{1}_i(g,1))}]\\
&=E[\frac{\mathbf{1}_i(g,1)\Delta Y_{i}}
{E(\mathbf{1}_i(g,1))}]
-
E[E[\frac{q_1(Z_{i}, \mathbf{X}, \mathbf{A})\mathbf{1}_i(g,0)\Delta Y_{i}}
{E(\mathbf{1}_i(g,1))}\,\big|\, \mathbf{U}, \mathbf{X}, \mathbf{A}]]\\
&=E[\frac{\mathbf{1}_i(g,1)\Delta Y_{i}}
{E(\mathbf{1}_i(g,1))}]-E[E[\frac{q_1(Z_{i}, \mathbf{X}, \mathbf{A})\Delta Y_{i}}
{E(\mathbf{1}_i(g,1))}\,\big|\, G_i=g, D_i =0, \mathbf{U}, \mathbf{X}, \mathbf{A}]
E[\mathbf{1}_i(g,0)|  \mathbf{U}, \mathbf{X}, \mathbf{A}]]\\
&=E[\frac{\mathbf{1}_i(g,1)\Delta Y_{i}}
{E(\mathbf{1}_i(g,1))}]-E[E[q_1(Z_{i}, \mathbf{X}, \mathbf{A})
\,\big|\, G_i=g, D_i =0, \mathbf{U}, \mathbf{X}, \mathbf{A}]
E[\frac{\mathbf{1}_i(g,0)\Delta Y_{i}}{E(\mathbf{1}_i(g,1))}|  \mathbf{U}, \mathbf{X}, \mathbf{A}]]\\
&=E[\frac{\mathbf{1}_i(g,1)\Delta Y_{i}}
{E(\mathbf{1}_i(g,1))}]
-E[E[q_1(Z_{i}, \mathbf{X}, \mathbf{A})
\,\big|\, G_i=g, D_i =0, \mathbf{U}, \mathbf{X}, \mathbf{A}]
\frac{\mathbf{1}_i(g,0)\Delta Y_{i}}{E(\mathbf{1}_i(g,1))}],
\end{split}
\end{equation}
where the third equality follows from Assumption 2.3(b), $Z_{i}\perp \Delta Y_{i} \mid D_{i},G_{i},\mathbf{U}, \mathbf{X}, \mathbf{A}$. 
\par
Under Assumption 2.6, $E[\mathbf{1}_i(g,1)|\mathbf{U}, \mathbf{X}, \mathbf{A}]/
E[\mathbf{1}_i(g,0)|  \mathbf{U}, \mathbf{X}, \mathbf{A}]
=
E[q_1( Z_{i}, \mathbf{X}, \mathbf{A})|G_{i}=g,D_{i}=0,  \mathbf{U}, \mathbf{X},\mathbf{A}]$, we get
\begin{equation}
\begin{split}
\nonumber
&\quad E[Y_{i1}(1,\mathbf{D_{-i}})-Y_{i1}(0,\mathbf{D_{-i}})
|G_{i}=g, D_{i}=1]=
E[\frac{\mathbf{1}_i(g,1)\Delta Y_{i}}
{E(\mathbf{1}_i(g,1))}]
-
E[\frac{q_1(Z_{i}, \mathbf{X}, \mathbf{A})\mathbf{1}_i(g,0)\Delta Y_{i}}
{E(\mathbf{1}_i(g,1))}].
\end{split}
\end{equation}
Next, we prove that the confounding bridge function is identified by
\begin{equation}
\begin{split}
\nonumber
\frac{E[\mathbf{1}_i(g,1)|W_i, \mathbf{X}, \mathbf{A}]}
{E[\mathbf{1}_i(g,0)|  W_i, \mathbf{X}, \mathbf{A}]}
=
E[q_1( Z_{i}, \mathbf{X}, \mathbf{A})|G_{i}=g,D_{i}=0, W_i, \mathbf{X},\mathbf{A}].
\end{split}
\tag{A.2}
\end{equation}
using Assumption 2.3, we have
\begin{equation}
\begin{split}
\nonumber
&\quad\frac{E[\mathbf{1}_i(g,1)|W_i, \mathbf{X}, \mathbf{A}]}
{E[\mathbf{1}_i(g,0)|  W_i, \mathbf{X}, \mathbf{A}]}
\\&=E\left[\frac{E[\mathbf{1}_i(g,1)|W_i, \mathbf{X}, \mathbf{A}]}
{E[\mathbf{1}_i(g,0)|  W_i, \mathbf{X}, \mathbf{A}]}\,\Bigg|\, W_i, \mathbf{X}, \mathbf{A}
\right]\\
&=E\left[\frac{E[\mathbf{1}_i(g,1)|W_i, \mathbf{U},\mathbf{X}, \mathbf{A}]\mathbf{1}_i(g,0)}
{E[\mathbf{1}_i(g,0)|  W_i, \mathbf{X}, \mathbf{A}]
E[\mathbf{1}_i(g,0)|  W_i, \mathbf{U},\mathbf{X}, \mathbf{A}]}\,\Bigg|\, W_i, \mathbf{X}, \mathbf{A}
\right]\\
&=E\left[\frac{E[\mathbf{1}_i(g,1)|W_i, \mathbf{U},\mathbf{X}, \mathbf{A}]}
{
E[\mathbf{1}_i(g,0)|  W_i, \mathbf{U},\mathbf{X}, \mathbf{A}]}\,\Bigg|\, G_{i}=g, D_{i}=0,W_i, \mathbf{X}, \mathbf{A}
\right]\\
&=E\left[\frac{E[\mathbf{1}_i(g,1)|\mathbf{U},\mathbf{X}, \mathbf{A}]}
{
E[\mathbf{1}_i(g,0)|  \mathbf{U},\mathbf{X}, \mathbf{A}]}\,\Bigg|\, G_{i}=g, D_{i}=0,W_i, \mathbf{X}, \mathbf{A}
\right].\\
\end{split}
\end{equation}
Similarly, we have 
\begin{equation}
\begin{split}
\nonumber
&\quad E[q_1( Z_{i}, \mathbf{X}, \mathbf{A})|G_{i}=g,D_{i}=0, W_i, \mathbf{X},\mathbf{A}]\\&
=E[E[q_1( Z_{i}, \mathbf{X}, \mathbf{A})|G_{i}=g,D_{i}=0, W_i, \mathbf{U},\mathbf{X},\mathbf{A}]
|G_{i}=g,D_{i}=0, W_i, \mathbf{X},\mathbf{A}]\\
&=E[E[q_1( Z_{i}, \mathbf{X}, \mathbf{A})|G_{i}=g,D_{i}=0, \mathbf{U},\mathbf{X},\mathbf{A}]
|G_{i}=g,D_{i}=0, W_i, \mathbf{X},\mathbf{A}].\\
\end{split}
\end{equation}
Then, under Assumption 2.6, we proof
\begin{equation}
\begin{split}
\nonumber
\frac{E[\mathbf{1}_i(g,1)|W_i, \mathbf{X}, \mathbf{A}]}
{E[\mathbf{1}_i(g,0)|  W_i, \mathbf{X}, \mathbf{A}]}
=
E[q_1( Z_{i}, \mathbf{X}, \mathbf{A})|G_{i}=g,D_{i}=0, W_i, \mathbf{X},\mathbf{A}].
\end{split}
\end{equation}
Finally, under Assumption 2.7, we show that the unique solution to equation (A.2) identifies the outcome confounding bridge function $h$
. Suppose there are two functions $q_1(Z_{i}, g, \mathbf{X}, \mathbf{A})$ and $q'_1(Z_{i}, g, \mathbf{X}, \mathbf{A})$ that satisfy equation (A.2). Then,
\begin{equation}
\begin{split}
\nonumber
E[q_1(Z_{i}, \mathbf{X}, \mathbf{A})
-q'_1(Z_{i}, \mathbf{X}, \mathbf{A})
|G_{i}=g, D_{i}=0, W_{i}=w, \mathbf{X}= \mathbf{x}, \mathbf{A}]=0
\end{split}
\end{equation}
for all g, $\mathbf{x}$, $\mathbf{A}$, and almost all $w$. Then, under Assumption 2.7, $q_1(Z_{i}, \mathbf{X}, \mathbf{A})$ = $q'_1(Z_{i}, \mathbf{X}, \mathbf{A})$ almost surely. Thus, the solution to equation (A.2) identifies the outcome confounding bridge function.

\subsection*{A.3 Proof of Proposition 1}
Under $\mathcal{M}_1$ where working model $h^*_1(W_{i}, \mathbf{X}, \mathbf{A})$ is correctly specified, we have $h^*_1(W_{i}, \mathbf{X}, \mathbf{A})$ = $h_1(W_{i}, \mathbf{X}, \mathbf{A})$ for $g$ $\in$ $\mathcal{G}$. We have
\begin{equation}
\begin{split}
\nonumber
\tau^{dr}_{ADT}(g)&=\frac{1}{n}\sum_{i \in N_{n}} 
E\left[\left(\frac{\mathbf{1}_i(g,1)}
{E(\mathbf{1}_i(g,1))}
-
\frac{q^*_1(Z_{i}, \mathbf{X}, \mathbf{A})\mathbf{1}_i(g,0)}
{E(\mathbf{1}_i(g,1))}\right)
\left(\Delta Y_{i}-h^*_1(W_{i}, \mathbf{X}, \mathbf{A})\right)\right]\\
&=\frac{1}{n}\sum_{i \in N_{n}} 
E\left[\left(\frac{\mathbf{1}_i(g,1)}
{E(\mathbf{1}_i(g,1))}
-
\frac{q^*_1(Z_{i}, \mathbf{X}, \mathbf{A})\mathbf{1}_i(g,0)}
{E(\mathbf{1}_i(g,1))}\right)
\left(\Delta Y_{i}-h_1(W_{i}, \mathbf{X}, \mathbf{A})\right)\right]\\
&=\frac{1}{n}\sum_{i \in N_{n}} 
E[\Delta Y_{i}-h_1(W_{i}, \mathbf{X}, \mathbf{A})\,|\, G_i=g, D_i=1]\\
&-\frac{1}{n}\sum_{i \in N_{n}}E\left[
\frac{q^*_1(Z_{i}, \mathbf{X}, \mathbf{A})}
{E(\mathbf{1}_i(g,1))}
E[\mathbf{1}_i(g,0)\Delta Y_{i}
-\mathbf{1}_i(g,0)h_1(W_{i}, \mathbf{X}, \mathbf{A})\,|\,Z_i, \mathbf{X}, \mathbf{A}]\right]\\
&=\frac{1}{n}\sum_{i \in N_{n}} 
E[\Delta Y_{i}-h_1(W_{i}, \mathbf{X}, \mathbf{A})\,|\, G_i=g, D_i=1]\\
&-\frac{1}{n}\sum_{i \in N_{n}}E\Bigg[
\frac{q^*_1(Z_{i}, \mathbf{X}, \mathbf{A})\mathbf{1}_i(g,0)}
{E(\mathbf{1}_i(g,1))}
(E[\Delta Y_{i}\,|\, G_i=g, D_i=0,Z_i, \mathbf{X}, \mathbf{A}]
\\&\quad\quad\quad\quad\quad\quad\quad\quad\quad\quad\quad\quad
-E[h_1(W_{i}, \mathbf{X}, \mathbf{A})\,|\, G_i=g, D_i=0,Z_i, \mathbf{X}, \mathbf{A}]
)
\Bigg]\\
&=\frac{1}{n}\sum_{i \in N_{n}} 
E[\Delta Y_{i}-h_1(W_{i}, \mathbf{X}, \mathbf{A})\,|\, G_i=g, D_i=1],\\
\end{split}
\end{equation}
where the third equation is obtained by  the law of iterated expectations, the fourth equation  follows from 
\begin{equation}
\begin{split}
\nonumber
&\quad E[E[\mathbf{1}_i(g,0)\Delta Y_i|Z_i, \mathbf{X}, \mathbf{A}]]
=E[E[\Delta Y_i|G_i=g,D_i=0,Z_i, \mathbf{X}, \mathbf{A}]\mathbf{1}_i(g,0)],
\end{split}
\end{equation}
and the last equation  follows from the equation (5) in Theorem 1. Based on the identification result of Theorem 1, we have $\tau^{dr}_{ADT}(g)$ = $\tau_{ADT}(g)$.
\par
Under $\mathcal{M}_2$ where working model $q^*_1(Z_{i}, \mathbf{X}, \mathbf{A})$ is correctly specified, we have $q^*_1(Z_{i}, \mathbf{X}, \mathbf{A})$ = $q_1(W_{i}, \mathbf{X}, \mathbf{A})$ for $g$ $\in$ $\mathcal{G}$. We have
\begin{equation}
\begin{split}
\nonumber
\tau^{dr}_{ADT}(g)&=\frac{1}{n}\sum_{i \in N_{n}} 
E\left[\left(\frac{\mathbf{1}_i(g,1)}
{E(\mathbf{1}_i(g,1))}
-
\frac{q^*_1(Z_{i}, \mathbf{X}, \mathbf{A})\mathbf{1}_i(g,0)}
{E(\mathbf{1}_i(g,1))}\right)
\left(\Delta Y_{i}-h^*_1(W_{i}, \mathbf{X}, \mathbf{A})\right)\right]\\
&=\frac{1}{n}\sum_{i \in N_{n}} 
E\left[\left(\frac{\mathbf{1}_i(g,1)}
{E(\mathbf{1}_i(g,1))}
-
\frac{q_1(Z_{i}, \mathbf{X}, \mathbf{A})\mathbf{1}_i(g,0)}
{E(\mathbf{1}_i(g,1))}\right)
\left(\Delta Y_{i}-h^*_1(W_{i},\mathbf{X}, \mathbf{A})\right)\right]\\
&=\frac{1}{n}\sum_{i \in N_{n}} 
E\left[\left(\frac{\mathbf{1}_i(g,1)}
{E(\mathbf{1}_i(g,1))}
-
\frac{q_1(Z_{i}, \mathbf{X}, \mathbf{A})\mathbf{1}_i(g,0)}
{E(\mathbf{1}_i(g,1))}\right)
\Delta Y_{i}\right]\\
&-\frac{1}{n}\sum_{i \in N_{n}} 
E\left[h^*_1(W_{i}, \mathbf{X}, \mathbf{A})E\left[\left(\frac{\mathbf{1}_i(g,1)}
{E(\mathbf{1}_i(g,1))}
-
\frac{q_1(Z_{i}, \mathbf{X}, \mathbf{A})\mathbf{1}_i(g,0)}
{E(\mathbf{1}_i(g,1))}\right)
\,|\,W_i, \mathbf{X}, \mathbf{A}\right]\right]\\
&=\frac{1}{n}\sum_{i \in N_{n}} 
E\left[\left(\frac{\mathbf{1}_i(g,1)}
{E(\mathbf{1}_i(g,1))}
-
\frac{q_1(Z_{i}, \mathbf{X}, \mathbf{A})\mathbf{1}_i(g,0)}
{E(\mathbf{1}_i(g,1))}\right)
\Delta Y_{i}\right]\\
&-\frac{1}{n}\sum_{i \in N_{n}} 
E\Bigg[\frac{h^*_1(W_{i}, \mathbf{X}, \mathbf{A})}{E(\mathbf{1}_i(g,1))}
(E[\mathbf{1}_i(g,1)\,|\,W_i, \mathbf{X}, \mathbf{A}]
\\&\quad\quad\quad\quad\quad\quad
-E[q_1(Z_{i}, \mathbf{X}, \mathbf{A})\,|\,G_i=g, D_i=0, W_i, \mathbf{X}, \mathbf{A}]E[\mathbf{1}_i(g,0)|  W_i, \mathbf{X}, \mathbf{A}]
\Bigg]\\
&=\frac{1}{n}\sum_{i \in N_{n}} 
E\left[\left(\frac{\mathbf{1}_i(g,1)}
{E(\mathbf{1}_i(g,1))}
-
\frac{q_1(Z_{i}, \mathbf{X}, \mathbf{A})\mathbf{1}_i(g,0)}
{E(\mathbf{1}_i(g,1))}\right)
\Delta Y_{i}\right]\\
&-\frac{1}{n}\sum_{i \in N_{n}} 
E\left[\frac{h^*_1(W_{i}, \mathbf{X}, \mathbf{A})}{E(\mathbf{1}_i(g,1))}
(E[\mathbf{1}_i(g,1)\,|\,W_i, \mathbf{X}, \mathbf{A}]
-\frac{E[\mathbf{1}_i(g,1)|W_i, \mathbf{X}, \mathbf{A}]}
{E[\mathbf{1}_i(g,0)|  W_i, \mathbf{X}, \mathbf{A}]}E[\mathbf{1}_i(g,0)|  W_i, \mathbf{X}, \mathbf{A}])
\right]\\
&=\frac{1}{n}\sum_{i \in N_{n}} 
E\left[\left(\frac{\mathbf{1}_i(g,1)}
{E(\mathbf{1}_i(g,1))}
-
\frac{q_1(Z_{i}, \mathbf{X}, \mathbf{A})\mathbf{1}_i(g,0)}
{E(\mathbf{1}_i(g,1))}\right)
\Delta Y_{i}\right],\\
\end{split}
\end{equation}
where the third equation is obtained by  the law of iterated expectations, the fourth equation  follows from 
\begin{equation}
\begin{split}
\nonumber
&\quad E[E[\mathbf{1}_i(g,0)q_1(Z_{i}, \mathbf{X}, \mathbf{A})|W_i, \mathbf{X}, \mathbf{A}]]
=E[E[q_1(Z_{i}, \mathbf{X}, \mathbf{A})|G_i=g,D_i=0,W_i, \mathbf{X}, \mathbf{A}]\mathbf{1}_i(g,0)],
\end{split}
\end{equation}
and the fifth equation  follows from the equation (6) in Theorem 2. Based on the identification result of Theorem 2, we have $\tau^{dr}_{ADT}(g)$ = $\tau_{ADT}(g)$. Together, we complete the proof.

\subsection*{A.4 Proof of Theorem 3}
This proof adopts the methodology of \cite{leung2022} and adapts it to our framework. Let $\mathcal{F}_n$ be the $\sigma$-algebra generated by $(\mathbf{Z}, \mathbf{X}, \mathbf{A})$, either $h, h' \in \mathbb{N}$, $f \in \mathcal{L}_h$, and $f' \in \mathcal{L}_{h'}$, or $h = h' = 1$ and
$f = f' = i^*$, $s > 0$ and $(H, H') \in P(h, h'; s)$. Define $\xi = f(\mathbf{C_H})$ and $\zeta = f'(\mathbf{C_{H'}})$.
Let $\mathbf{D'}, \mathbf{D''}$ each be independent copies of $\mathbf{D}$. Define $\mathbf{D_i^{(s,\xi)}}$ = $(\mathbf{D}_{N_{\mathbf{A}}(i,s)}, \mathbf{D}'_{N^c_{\mathbf{A}}(i,s)})$, $\mathbf{D_i^{(s,\zeta)}}$ = $(\mathbf{D}_{N_{\mathbf{A}}(i,s)}, \mathbf{D}''_{N^c_{\mathbf{A}}(i,s)})$, $\mathbf{D_{-i}^{(s,\xi)}}$ = $(\mathbf{D}_{N_{\mathbf{A}}(i,s)}, \mathbf{D}'_{N^c_{\mathbf{A}}(i,s)})\texttt{\textbackslash}D_i$, $\mathbf{D_{-i}^{(s,\zeta)}}$ = $(\mathbf{D}_{N_{\mathbf{A}}(i,s)}, \mathbf{D}''_{N^c_{\mathbf{A}}(i,s)})\texttt{\textbackslash}D_i$, and 
\begin{equation}
\begin{split}
\nonumber
C_i^{(s,\xi)} = \left(\frac{\mathbf{1}^{(s,\xi)}_i(g,1)}
{E(\mathbf{1}_i(g,1))}
-
\frac{q^*_1(Z_{i}, \mathbf{X}, \mathbf{A})\mathbf{1}^{(s,\xi)}_i(g,0)}
{E(\mathbf{1}_i(g,1))}\right)
\left(\Delta Y_{i}(\mathbf{D_{i}^{(s,\xi)}})-h^*_1(W_{i}, \mathbf{X}, \mathbf{A})\right),
\end{split}
\end{equation}
where $\mathbf{1}^{(s,\xi)}_i(g,d)$ = $\mathbf{1}\{G(i, \mathbf{D_{-i}^{(s,\xi)}}, \mathbf{A})=g, D_i=d\}$ for $d$ = 0, 1. The definition of $C_i^{(s,\zeta)}$ is similar to that of $C_i^{(s,\xi)}$. Finally let $\xi^{(s)}$ = $f((C_i^{(s,\xi)} : i \in H))$ and $\zeta^{(s)}$ = $f((C_i^{(s,\zeta)} : i \in H'))$. Based on Assumptions 3.2 and 3.3, $C_i$ is uniformly bounded, so $|Cov(\xi,\zeta|\mathbf{Z}, \mathbf{X}, \mathbf{A})| \le 2||f||_{\infty}||f'||_{\infty}$, so for $s$ $\le$ 2max$\{K,1\}$, we have $|Cov(\xi,\zeta|\mathbf{Z}, \mathbf{X}, \mathbf{A})| \le  \Psi_{h,h'} (f ,f')$. Now consider $s$ > 2max$\{K,1\}$, so that $\ell_{\mathbf{A}}(H,H')$ > 2max$\{K,1\}$. By Assumption 3.4, $(C_i^{(\lfloor s/2 \rfloor,\xi)} : i \in H)$ $\perp$ $(C_i^{(\lfloor s/2 \rfloor,\zeta)} : i \in H') | \mathbf{Z}, \mathbf{X}, \mathbf{A}$. Then
\begin{equation}
\begin{split}
\nonumber
|Cov(\xi,\zeta|\mathbf{Z}, \mathbf{X}, \mathbf{A})|&\le |Cov(\xi-\xi^{(\lfloor s/2 \rfloor)},\zeta | \mathbf{Z}, \mathbf{X}, \mathbf{A})|
+|Cov(\xi^{(\lfloor s/2 \rfloor| \mathbf{Z}, \mathbf{X}, \mathbf{A})},\zeta-\zeta^{(\lfloor s/2 \rfloor)}| \mathbf{Z}, \mathbf{X}, \mathbf{A})|\\
&\le 2||f'||_{\infty}E[|\xi-\xi^{(\lfloor s/2 \rfloor)}\,\big|\, \mathbf{Z}, \mathbf{X}, \mathbf{A}]
+2||f||_{\infty}E[|\zeta-\zeta^{(\lfloor s/2 \rfloor)}|\,\big|\, \mathbf{Z}, \mathbf{X}, \mathbf{A}]\\
&\le 2(h||f'||_{\infty}Lip(f)+h'||f||_{\infty}Lip(f'))\theta^{ADT}_{n,\lfloor s/2 \rfloor},
\end{split}
\end{equation}
where the last line uses the fact that, by Assumption 3.4 (b), $\mathbf{1}^{(s,\xi)}_i(g,d)$ = $\mathbf{1}_i(g,d)$ for $d$ = 0, 1, and by Assumption 3.5, $\max_{i\in N_n}E[|\Delta Y_{i}(\mathbf{D})-\Delta Y_{i}(\mathbf{D}_i^{(\lfloor s/2 \rfloor,\xi)})|\,\big|\, \mathbf{Z}, \mathbf{X}, \mathbf{A}]\le \theta^{ADT}_{n,\lfloor s/2 \rfloor}$. 

\subsection*{A.5 Proof of Theorem 4}
Decompose
\begin{equation}
\begin{split}
\nonumber
\sqrt{n}(\hat{\tau}^{dr}_{ADT}(g)-\tau^{dr}_{ADT}(g))
&=\frac{1}{\sqrt{n}}\sum_{i\in N_n}(C_i-\tau^{mr}_{ADT}(g))
+C_{i1}+C_{i2}+C_{i3}+C_{i4}+C_{i5}+C_{i6}+C_{i7},\\
\end{split}
\end{equation} 
where
\begin{equation}
\begin{split}
\nonumber
C_{i1}=\frac{1}{\sqrt{n}}\sum_{i \in N_{n}} 
\frac{
(q^*_1(Z_{i}, \mathbf{X}, \mathbf{A})-\hat{q}^*_1(Z_{i}, \mathbf{X}, \mathbf{A}))\mathbf{1}_i(g,0)}
{E(\mathbf{1}_i(g,1))}
\left(\Delta Y_{i}-h^*_1(W_{i}, \mathbf{X}, \mathbf{A})\right),
\end{split}
\end{equation}
\begin{equation}
\begin{split}
\nonumber
C_{i2}=\frac{1}{\sqrt{n}}\sum_{i \in N_{n}} 
\frac{\mathbf{1}_i(g,1)-
q^*_1(Z_{i}, \mathbf{X}, \mathbf{A})\mathbf{1}_i(g,0)}
{E(\mathbf{1}_i(g,1))}
\left(h^*_1(W_{i}, \mathbf{X}, \mathbf{A})-\hat{h}^*_1(W_{i}, \mathbf{X}, \mathbf{A})\right),
\end{split}
\end{equation}
\begin{equation}
\begin{split}
\nonumber
C_{i3}=\frac{1}{\sqrt{n}}\sum_{i \in N_{n}} 
\frac{
(q^*_1(Z_{i}, \mathbf{X}, \mathbf{A})-\hat{q}^*_1(Z_{i}, \mathbf{X}, \mathbf{A}))\mathbf{1}_i(g,0)}
{E(\mathbf{1}_i(g,1))}
\left(h^*_1(W_{i}, \mathbf{X}, \mathbf{A})-\hat{h}^*_1(W_{i}, \mathbf{X}, \mathbf{A})\right),
\end{split}
\end{equation}
\begin{equation}
\begin{split}
\nonumber
C_{i4}=\frac{1}{\sqrt{n}}\sum_{i \in N_{n}} 
\frac{\mathbf{1}_i(g,1)-
q^*_1(Z_{i}, \mathbf{X}, \mathbf{A})\mathbf{1}_i(g,0)}
{E(\mathbf{1}_i(g,1))\hat{E}(\mathbf{1}_i(g,1))}
\left(\Delta Y_{i}-h^*_1(W_{i}, \mathbf{X}, \mathbf{A})\right)
\left(E(\mathbf{1}_i(g,1))-\hat{E}(\mathbf{1}_i(g,1))\right),
\end{split}
\end{equation}
\begin{equation}
\begin{split}
\nonumber
C_{i5}&=\frac{1}{\sqrt{n}}\sum_{i \in N_{n}} 
\frac{
(q^*_1(Z_{i}, \mathbf{X}, \mathbf{A})-\hat{q}^*_1(Z_{i}, \mathbf{X}, \mathbf{A}))\mathbf{1}_i(g,0)}
{E(\mathbf{1}_i(g,1))\hat{E}(\mathbf{1}_i(g,1))}
\left(\Delta Y_{i}-h^*_1(W_{i}, \mathbf{X}, \mathbf{A})\right)\\&
\quad \times
\left(E(\mathbf{1}_i(g,1))-\hat{E}(\mathbf{1}_i(g,1))\right),
\end{split}
\end{equation}
\begin{equation}
\begin{split}
\nonumber
C_{i6}&=\frac{1}{\sqrt{n}}\sum_{i \in N_{n}} 
\frac{\mathbf{1}_i(g,1)-
q^*_1(Z_{i}, \mathbf{X}, \mathbf{A})\mathbf{1}_i(g,0)}
{E(\mathbf{1}_i(g,1))\hat{E}(\mathbf{1}_i(g,1))}
\left(h^*_1(W_{i}, \mathbf{X}, \mathbf{A})-\hat{h}^*_1(W_{i}, \mathbf{X}, \mathbf{A})\right)\\&\quad\times
\left(E(\mathbf{1}_i(g,1))-\hat{E}(\mathbf{1}_i(g,1))\right),
\end{split}
\end{equation}
\begin{equation}
\begin{split}
\nonumber
C_{i7}&=\frac{1}{\sqrt{n}}\sum_{i \in N_{n}} 
\frac{
(q^*_1(Z_{i}, \mathbf{X}, \mathbf{A})-\hat{q}^*_1(Z_{i}, \mathbf{X}, \mathbf{A}))\mathbf{1}_i(g,0)}
{E(\mathbf{1}_i(g,1))\hat{E}(\mathbf{1}_i(g,1))}
\left(h^*_1(W_{i}, \mathbf{X}, \mathbf{A})-\hat{h}^*_1(W_{i}, \mathbf{X}, \mathbf{A})\right)\\&\quad\times
\left(E(\mathbf{1}_i(g,1))-\hat{E}(\mathbf{1}_i(g,1))\right).
\end{split}
\end{equation}
\par
For $C_{i1}$, there exist universal constants C > 0,  
\begin{equation}
\begin{split}
\nonumber
\mathrm{E}((C_{i1})^2) \le
&\frac{C}{n}\sum_{i \in N_{n}} \sum_{j \in N_{n}}\mathrm{E}\Bigg[
\mathrm{E}\Bigg[(\mathbf{1}_i(g,0) (\Delta Y_i - h^*_1(W_{i}, \mathbf{X}, \mathbf{A}))
\mathbf{1}_j(g,0) (\Delta Y_j - h^*_1(W_{j}, \mathbf{X}, \mathbf{A}))
\,\Big|\,\mathbf{Z},\mathbf{X}, \mathbf{A}\Bigg]
\\
&\quad \times
   (\hat{q}^*_1(Z_{i}, \mathbf{X}, \mathbf{A})-q^*_1(Z_{i}, \mathbf{X}, \mathbf{A}))
   (\hat{q}^*_1(Z_{j}, \mathbf{X}, \mathbf{A})-q^*_1(Z_{j}, \mathbf{X}, \mathbf{A}))
 \Bigg]\\
&\le C\sum_{s=0}^{n-1}\Psi_{h,h'}(f,f') \, (\widetilde{\theta}^{ADT}_{n,s})^{1-\varepsilon}
\mathrm{E}\Bigg[\frac{1}{n}\sum_{i \in N_{n}}
 (\hat{q}^*_1(Z_{i}, \mathbf{X}, \mathbf{A})-q^*_1(Z_{i}, \mathbf{X}, \mathbf{A})) 
\\&\quad\times           
\sum_{j \in N_{n}}\mathbf{1}\{\ell_{\mathbf{A}}(i,j)=s\}
   (\hat{q}^*_1(Z_{j}, \mathbf{X}, \mathbf{A})-q^*_1(Z_{j}, \mathbf{X}, \mathbf{A}))
\Bigg]\\
&\le C\sum_{s=0}^{n-1}\Psi_{h,h'}(f,f')
 \, (\widetilde{\theta}^{ADT}_{n,s})^{1-\varepsilon}M^{\partial}_{N_n}(s,2)^{\frac{1}{2}}
\sqrt{\mathrm{E}\Bigg[\frac{1}{n}\sum_{i \in N_{n}}
 (\hat{q}^*_1(Z_{i}, \mathbf{X}, \mathbf{A})-q^*_1(Z_{i}, \mathbf{X}, \mathbf{A})
)^2\Bigg]},
\end{split}
\end{equation}
where  the first inequality uses Assumption 3.3, the second inequality uses Lemma B.1, and  the last inequality uses Cauchy-Schwarz inequality. The last line is $o_P$(1) by Assumption 3.1(a) and 3.6 (b). 
\par
For $C_{i2}$, we directly have that $C_{i2}$ = $o_P(1)$ by Assumption 3.1 (c).  
\par
For $C_{i3}$, 
\begin{equation}
\begin{split}
\nonumber
|C_{i3}|\le\sqrt{n}\frac{1}{C}
\sqrt{\frac{1}{n}\sum_{i \in N_{n}}
(\hat{q}^*_1(Z_{i}, \mathbf{X}, \mathbf{A})-q^*_1(Z_{i}, \mathbf{X}, \mathbf{A}))^2 
\frac{1}{n}\sum_{i \in N_{n}}
(\hat{h}^*_1(W_{i}, \mathbf{X}, \mathbf{A})-h^*_1(W_{i}, \mathbf{X}, \mathbf{A}))^2
}=o_P(1),
\end{split}
\end{equation}
by H$\ddot{\text{o}}$lder's inequality, Assumption 3.1 (b) and Assumption 3.3. 
\par
The proof of $C_{i4}$ and $C_{i5}$ are similar with $C_{i1}$. For $C_{i4}$, we have 
\begin{equation}
\begin{split}
\nonumber
\mathrm{E}((C_{i4})^2) 
&\le C\sum_{s=0}^{n-1}\Psi_{h,h'}(f,f') \, (\widetilde{\theta}^{ADT}_{n,s})^{1-\varepsilon}
\left(\hat{E}(\mathbf{1}_i(g,1))-E(\mathbf{1}_i(g,1))\right)^2\\
&=O_P(\frac{1}{n}),
\end{split}
\end{equation}
by Assumption 3.3, Assumption 3.6 (b), Lemma B.1 and Lemma B.2. 
\par
For $C_{i5}$, 
\begin{equation}
\begin{split}
\nonumber
\mathrm{E}((C_{i5})^2) 
&\le C\sum_{s=0}^{n-1}\Psi_{h,h'}(f,f') \, (\widetilde{\theta}^{ADT}_{n,s})^{1-\varepsilon}
\sqrt{\mathrm{E}\Bigg[\frac{1}{n}\sum_{i \in N_{n}}
(\hat{q}^*_1(Z_{i}, \mathbf{X}, \mathbf{A})-q^*_1(Z_{i}, \mathbf{X}, \mathbf{A})
)^2\Bigg]}
 \\&\quad \times
\left(\hat{E}(\mathbf{1}_i(g,1))-E(\mathbf{1}_i(g,1))\right)^2\\
&=o_P(1)
\end{split}
\end{equation}
by Assumption 3.1 (a), Assumption 3.3, Assumption 3.6 (b), Lemma B.1 and Lemma B.2. 
\par
The proof of $C_{i6}$ and $C_{i7}$ are similar with $C_{i3}$. For $C_{i6}$, 
\begin{equation}
\begin{split}
\nonumber
|C_{i6}|\le\sqrt{n}\frac{1}{C}
\sqrt{\frac{1}{n}\sum_{i \in N_{n}}
(\hat{q}^*_1(Z_{i}, \mathbf{X}, \mathbf{A})-q^*_1(Z_{i}, \mathbf{X}, \mathbf{A}))^2 
}\times|\hat{E}(\mathbf{1}_i(g,1))-E(\mathbf{1}_i(g,1))|=o_P(1),
\end{split}
\end{equation}
by H$\ddot{\text{o}}$lder's inequality, Assumption 3.1 (a), Assumption 3.3 and Lemma B.2. 
\par
For $C_{i7}$, 
\begin{equation}
\begin{split}
\nonumber
|C_{i7}|&\le\sqrt{n}\frac{1}{C}
\sqrt{\frac{1}{n}\sum_{i \in N_{n}}
(\hat{q}^*_1(Z_{i}, \mathbf{X}, \mathbf{A})-q^*_1(Z_{i}, \mathbf{X}, \mathbf{A}))^2 
\frac{1}{n}\sum_{i \in N_{n}}
(\hat{h}^*_1(W_{i}, \mathbf{X}, \mathbf{A})-h^*_1(W_{i}, \mathbf{X}, \mathbf{A}))^2
}\\&
\times|\hat{E}(\mathbf{1}_i(g,1))-E(\mathbf{1}_i(g,1))|\\
&=o_P(1),
\end{split}
\end{equation}
by H$\ddot{\text{o}}$lder's inequality, Assumption 3.1 (b), Assumption 3.3 and Lemma B.2. 
\par
 Totally, we have
\begin{equation}
\begin{split}
\nonumber
\sqrt{n}(\hat{\tau}^{dr}_{ADT}(g)-\tau^{dr}_{ADT}(g))
&=\frac{1}{\sqrt{n}}\sum_{i\in N_n}
\left(C_i-\tau^{dr}_{ADT}(g)\right)
+o_P(1).
\end{split}
\end{equation} 
By Theorem 2, $\{C_i\}_{i=1}^n$ is conditionally $\Psi$-dependence given $(\mathbf{Z}, \mathbf{X}, \mathbf{A})$ with the dependence coefficients $\{\widetilde{\theta}^{ADT}_{n,s}\}_{s \ge 0}$. Then, letting $\tilde{C}^{ADT}_{N_n}$ = $n^{-1/2}\sum_{i\in N_n}\left(C_i-\tau^{mr}_{ADT}(g)\right)/\sigma_n^{ADT}$, the same arguments as in the proofs of Lemmas A.2 and A.3 of Kojevnikov et al. (2021) show that there exists a positive constant C > 0 such that
\begin{equation}
\begin{split}
\nonumber
&\text{sup}_{a\in \mathbb{R}}\mid
 P(\tilde{C}^{ADT}_{N_n}\le a\mid \mathbf{Z},\mathbf{X}, \mathbf{A}-\phi(a))\mid\\
&\le\sum_{k=1}^2
\left(\sqrt{n^{-k/2}(\sigma_{N_n}^{ADT})^{-(2+k)}\sum_{s=0}^{n-1}
c_{N_n}(s,m_n;k)(\widetilde{\theta}^{ADT}_{n,s})^{1-\varepsilon}}
+n^{k/2}(\sigma_{N_n}^{ADT})^{-k}(\widetilde{\theta}^{ADT}_{n,m_n})^{1-\varepsilon}
\right),
\end{split}
\end{equation} 
where $\phi$ denotes the cumulative distribution function of $\mathcal{N}(0,1)$, $m_n$ and $\varepsilon$ are as given in Assumption 3.6.  The right-hand side converges to zero by Assumption 3.6, implying that $\tilde{C}^{ADT}_{N_n}$ $\overset{d}{\rightarrow} \mathcal{N}$(0, 1). Thus, we have
\begin{equation}
\begin{split}
\nonumber
(\sigma_n^{ADT})^{-1}\sqrt{n}(\hat{\tau}^{dr}_{ADT}(g)-\tau^{dr}_{ADT}(g))
=\tilde{C}^{ADT}_{N_n}+o_P(1)
\overset{d}{\rightarrow} \mathcal{N}(0, 1).
\end{split}
\end{equation} 

\subsection*{A.6 Proof of Theorem 5}
This proof is similar with \cite{leunggnn}. Define
\begin{equation}
\begin{split}
\nonumber
\tilde{\sigma}_n^{ADT}
 = \frac{1}{n}\sum_{i \in N_n}\sum_{j \in N_n}
\tau_{ADT,i}(g)\tau_{ADT,j}(g) \mathbf{1}\{\ell_{\mathbf{A}}(i,j) \le b_n\},
\end{split}
\end{equation} 
where
$\tau_{ADT,i}(g)$ = $C_i-\tau^{dr}_{ADT}(g)$. We first show that |$\hat{\sigma}_n^{ADT}$ - $\tilde{\sigma}_n^{ADT}$| $\overset{p}{\rightarrow} $ 0. We have
 \begin{equation}
\begin{split}
\nonumber
&\left |   \hat{\sigma}_n^{ADT} - \tilde{\sigma}_n^{ADT}\right |
=\left |  \frac{1}{n}\sum_{i \in N_n}(\hat{\tau}_{ADT,i}(g)-\tau_{ADT,i}(g))
\sum_{j\in N_n}(\hat{\tau}_{ADT,j}(g)+\tau_{ADT,j}(g))
\mathbf{1}\{\ell_{\mathbf{A}}(i,j) \le b_n\}
\right |\\
&\le \sqrt{\frac{1}{n}\sum_{i \in N_n}(\hat{\tau}_{ADT,i}(g)-\tau_{ADT,i}(g))^2}
\sqrt{\frac{1}{n}\sum_{i \in N_n}\text{max}_{j \in N_n}(\hat{\tau}_{ADT,j}(g)+\tau_{ADT,j}(g))^2
\left |N_{\mathbf{A}}(i,b_n)\right |^2}
\end{split}
\end{equation} 
   by H$\ddot{\text{o}}$lder's inequality. 
   \par
   Next, for some constant C > 0, 
 \begin{equation}
\begin{split}
\nonumber
\frac{1}{n}\sum_{i \in N_n}\text{max}_{j \in N_n}(\hat{\tau}_{ADT,j}(g)+\tau_{ADT,j}(g))^2
\left |N_{\mathbf{A}}(i,b_n)\right |^2
\le C\frac{1}{n}\sum_{i \in N_n}
\left |N_{\mathbf{A}}(i,b_n)\right |^2 = O_P(\sqrt{n}),
\end{split}
\end{equation} 
by  Assumptions 3.2, 3.3, 3.7(a) and (d). 
\par
Then, by Assumptions 3.2, 3.3, 3.7(a) and (b), we have
\begin{equation}
\begin{split}
\nonumber
\frac{1}{n}\sum_{i \in N_n}(\hat{\tau}_{ADT,i}(g)-\tau_{ADT,i}(g))^2
= o_P(n^{-1/2}). 
\end{split}
\end{equation} 
\par
Next, the proof of Theorem 4 of  \cite{leunggnn} can be applied to show that 
\begin{equation}
\begin{split}
\nonumber
\hat{\sigma}_n^{ADT} = \hat{\sigma}_n^{ADT*}+B_n+o_P(1).
\end{split}
\end{equation} 
The argument follows from substituting $\tilde{\tau}_{ADT,i}(g)$ for $Z_i$ - $\tau_i(t,t')$ in Theorem 4 of  \cite{leunggnn} and our Assumptions 3.7(c)-(e) for Assumptions 7(b)-(d). Finally, we apply Proposition 4.1 of \cite{kojevnikov2021limit} to show  $|\hat{\sigma}_n^{ADT*}-\sigma_n^{ADT}|
\overset{p}{\rightarrow}$ 0. 

\section*{B Lemmas}
$\mathbf{Lemma\ B.1.}$ Under Assumptions 3.2-3.5, 
\begin{equation}
\begin{aligned}
\nonumber
Cov(\mathbf{1}_i(g,d)(\Delta Y_i - \mu^{Y*}_{gi}(0,Z_{i}, \mathbf{X}, \mathbf{A})),\mathbf{1}_j(g,d)(\Delta Y_j - \mu^{Y*}_{gj}(0,Z_{i}, \mathbf{X}, \mathbf{A})) \,&\Big|\, \mathbf{Z}, \mathbf{X}, \mathbf{A}) \\&\le \Psi_{h,h'}(f,f') \, (\widetilde{\theta}^{ADT}_{n,s})^{1-\varepsilon},
\end{aligned}
\end{equation}
\begin{equation}
\begin{aligned}
\nonumber
Cov(\mathbf{1}_i(g,d)(W_i - \mu^{W*}_{gi}(d,Z_{i}, \mathbf{X}, \mathbf{A})),\mathbf{1}_j(g,d)(W_j - \mu^{W*}_{gj}(d,Z_{i}, \mathbf{X}, \mathbf{A})) \,&\Big|\, \mathbf{Z}, \mathbf{X}, \mathbf{A}) \\&\le \Psi_{h,h'}(f,f') \, (\widetilde{\theta}^{ADT}_{n,s})^{1-\varepsilon},
\end{aligned}
\end{equation}
for $d$ = 0, 1, a constant 0 < $\varepsilon$ < 1, holds with $\widetilde{\theta}^{ADT}_{n,s}$ = $\theta^{ADT}_{n,\lfloor s/2 \rfloor}\mathbf{1}\{s > 2max\{K,1\}\}+\mathbf{1}\{s \le 2max\{K,1\}\}$ for all $n$ $\in$ $\mathbb{N}$ and $s$ > 0 and 
\begin{equation}
\begin{split}
\nonumber
 \Psi_{h,h'} (f ,f')=2(||f||_{\infty}||f'||_{\infty}+h||f'||_{\infty}Lip(f)+h'||f||_{\infty}Lip(f'))
\end{split}
\end{equation} 
for  either $h$, $h'$ $\in$ $\mathbb{N}$, $f$ $\in$ $\mathcal{L}_h$, $f'$ $\in$ $\mathcal{L}_{h'}$, or $h$ = $h'$ =1 and $f$ = $f'$ = $i^*$. 
\par
$\mathbf{Proof:}$ Similar with the proof of Theorem 1, we have 
\begin{equation}
\begin{aligned}
\nonumber
Cov(\mathbf{1}_i(g,d)(\Delta Y_i - \mu^{Y*}_{gi}(0,Z_{i}, \mathbf{X}, \mathbf{A})),\mathbf{1}_j(g,d)(\Delta Y_j - \mu^{Y*}_{gj}(0,Z_{i}, \mathbf{X}, \mathbf{A})) \,&\Big|\, \mathbf{Z}, \mathbf{X}, \mathbf{A}) \\&\le \Psi_{h,h'}(f,f') \, \widetilde{\theta}^{ADT}_{n,s},
\end{aligned}
\end{equation}
\begin{equation}
\begin{aligned}
\nonumber
Cov(\mathbf{1}_i(g,d)(W_i - \mu^{W*}_{gi}(d,Z_{i}, \mathbf{X}, \mathbf{A})),\mathbf{1}_j(g,d)(W_j - \mu^{W*}_{gj}(d,Z_{i}, \mathbf{X}, \mathbf{A})) \,&\Big|\, \mathbf{Z}, \mathbf{X}, \mathbf{A}) \\&\le \Psi_{h,h'}(f,f') \, \widetilde{\theta}^{ADT}_{n,s},
\end{aligned}
\end{equation}
then, the proof was completed using Corollary A.2 of \cite{kojevnikov2021limit}.
\par
$\mathbf{Lemma\ B.2.}$ Suppose that Assumptions 3.2, 3.4 and 3.6  hold. Then, we have
\begin{equation}
\begin{split}
\nonumber
\hat{\mathrm{E}}[\mathbf{1}_i(g,d)]-\mathrm{E}[\mathbf{1}_i(g,d)]=O_P(\frac{1}{\sqrt{n}})
\end{split}
\end{equation} 
for all $g$ $\in$ $\mathcal{G}$ and $d$ = 0, 1. 
\par
$\mathbf{Proof:}$ By Assumptions 3.4 (c), we have $\mathrm{E}[\hat{\mathrm{E}}[\mathbf{1}_i(g,d)]]$ = $\mathrm{E}[\mathbf{1}_i(g,d)]$, and thus it suffices to show that $Var(\hat{\mathrm{E}}[\mathbf{1}_i(g,d)])$ = $O(\frac{1}{\sqrt{n}})$. Observe that
\begin{equation}
\begin{split}
\nonumber
Var(\hat{\mathrm{E}}[\mathbf{1}_i(g,d)])&=
\frac{1}{n^2}\sum_{i \in N_{n}}Var(\mathbf{1}_i(g,d))+
\frac{1}{n^2}\sum_{i \in N_{n}}\sum_{j \in N_{n}\setminus \{i\}}
Cov(\mathbf{1}_i(g,d),\mathbf{1}_j(g,d))\\
&=O(\frac{1}{n})+\frac{1}{n^2}\sum_{i \in N_{n}}\sum_{j \in N_{n}}
\sum_{s \ge 1}\mathbf{1}\{\ell_{\mathbf{A}}(i,j)=s\}
Cov(\mathbf{1}_i(g,d),\mathbf{1}_j(g,d))\\
&=O(\frac{1}{n})+\frac{1}{n^2}\sum_{i \in N_{n}}\sum_{j \in N_{n}}
\sum_{s = 1}^{2K}\mathbf{1}\{\ell_{\mathbf{A}}(i,j)=s\}
Cov(\mathbf{1}_i(g,d),\mathbf{1}_j(g,d)),\\
\end{split}
\end{equation} 
where the last equality follows from Assumption 3.4. By the Cauchy-Schwarz inequality, the second term of the last line is bounded above by $n^{-1}\sum_{s = 1}^{2K}M^{\partial}_{N_n}(s)$ which is $O(n^{-1})$ by Assumption 3.6 (a).

\section*{C Average indirect effect}
The proof of AIT follows a similar approach to that of ADT. Therefore, in the following, we present the propositions and theorems related to AIT and omit the proof. First, we propose the following identifying assumption for AIT:
\par
$\mathbf{Assumption\ C.1.}$
\\
(a) (Latent parallel trends for AIT)
\begin{equation}
\begin{split}
\nonumber
&\quad\frac{1}{n}\sum_{i \in N_{n}}
E[\sum_{j \in \mathcal{E} _{i}}(Y_{j1}(0,\mathbf{D_{-i}})-Y_{j0}(0,\mathbf{D_{-i}}))|D_{i}=1, \mathbf{U}, \mathbf{X}, \mathbf{A}]\\
&=\frac{1}{n}\sum_{i \in N_{n}}
E[\sum_{j \in \mathcal{E} _{i}}(Y_{j1}(0,\mathbf{D_{-i}})-Y_{j0}(0,\mathbf{D_{-i}}))|D_{i}=0, \mathbf{U}, \mathbf{X},\mathbf{A}]. 
\end{split}
\end{equation}
(b) (Negative controls)
\\
Negative control outcome (NCO): For all $i$ $\in$ $N_{n}$ and all $j$ $\in$ $\mathcal{E} _{i}$, $W_{j}$ satisfy
\begin{equation}
\nonumber
 W_{j}\perp D_{i} \mid \mathbf{U}, \mathbf{X}, \mathbf{A}.
\end{equation}
Negative control exposure (NCE): For all $i$ $\in$ $N_{n}$ and all $j$ $\in$ $\mathcal{E} _{i}$, $Z_{i}$ satisfy
\begin{equation}
\begin{split}
\nonumber
&Z_{i}\perp \Delta Y_{j} \mid D_{i}, \mathbf{U}, \mathbf{X}, \mathbf{A},\quad \text{and}
\\&Z_{i}\perp  W_{j} \mid D_{i}, \mathbf{U}, \mathbf{X}, \mathbf{A}.
\end{split}
\end{equation}
\par
$\mathbf{Assumption\ C.2.}$
\\
(a) (Outcome confounding bridge function)
There exists a function $h_2$($W_{j}$, $D_{i}$, $\mathbf{X}$, $\mathbf{A}$), such that for all $i$ $\in$ $N_{n}$ and all $j$ $\in$ $\mathcal{E} _{i}$,
\begin{equation}
\begin{split}
\nonumber
E[\Delta Y_{j}|D_{i}=0, \mathbf{U}, \mathbf{X}, \mathbf{A}]
=
E[h_2( W_{j},  \mathbf{X}, \mathbf{A})|D_{i}=0, \mathbf{U}, \mathbf{X},\mathbf{A}]. 
\end{split}
\end{equation}
(b) (Negative control relevance)
For any square integrable function $f$ and any $d$, $\mathbf{x}$ and $\mathbf{A}$, if E($f(W_{j}$) $\mid$ $Z_{i}$ =$z$, $D_{i}$ = $0$, $\mathbf{X}$ = $\mathbf{x}$, $\mathbf{A}$) = 0 for almost all $z$, then $f(W_{j}$) = 0 almost surely.
\par
Then, we establish non-parametric identification of the AIT under 
Assumption 2.2, C.1 and C.2. 
\par
$\mathbf{Theorem\ C.1}$
Under Assumption 2.2, C.1 and C.2, the confounding bridge function is identified as the unique solution to the following equation:
\begin{equation}
\begin{split}
\nonumber
E[\Delta Y_{j}|Z_{i}, D_{i}=0, \mathbf{X}, \mathbf{A}]
=
E[h_2(W_{j},  \mathbf{X}, \mathbf{A})|Z_{i}, D_{i}=0, \mathbf{X}, \mathbf{A}], 
\end{split}
\end{equation}
and the AIT is identified by
\begin{equation}
\begin{split}
\nonumber
\tau_{AIT}=\frac{1}{n}\sum_{i \in N_{n}}
E[\sum_{j \in \mathcal{E} _{i}}(\Delta Y_j
-h_2(W_{j}, \mathbf{X}, \mathbf{A}))
|D_{i}=1].
\end{split}
\end{equation}
In a DAG framework similar to Figure 2, given a sufficiently large sample size, $D_{j}$ satisfies the NCO conditions, while the treatments of units located at a distance of at least 2 from individual $j$ $\in \mathcal{E} _{i}$ fulfill the NCE requirements. 
\par
Second, we givean alternative identification approach and subsequently derive the doubly robust DID estimands for AIT. 

$\mathbf{Assumption\ C.3.}$
\\
(a) (Treatment confounding bridge function) There exists a function $q_2$($W_{j}$, $\mathbf{X}$, $\mathbf{A}$), such that for all $i$ $\in$ $N_{n}$ and all $j$ $\in$ $\mathcal{E} _{i}$,
\begin{equation}
\begin{split}
\nonumber
\frac{E[D_i\mid \mathbf{U}, \mathbf{X}, \mathbf{A}]}
{E[1-D_i\mid  \mathbf{U}, \mathbf{X}, \mathbf{A}]}
=
E[q_2( Z_{j}, \mathbf{X}, \mathbf{A})\mid D_{i}=0,  \mathbf{U}, \mathbf{X},\mathbf{A}].
\end{split}
\end{equation}
(b) (Negative control relevance) For any square integrable function $f$, $\mathbf{x}$ and $\mathbf{A}$, if E(f($Z_{j}$) $\mid$ $D_{i}$ = $0$, $W_{i}$ =$w$, $\mathbf{X}$ = $\mathbf{x}$, $\mathbf{A}$) = 0 for almost all $w$, then f($Z_{i}$) = 0 almost surely.
\par
we establish another identification of the AIT under Assumption 2.2, C.1 and C.3.
$\mathbf{Theorem\ C.2}$
Under Assumption 2.2, C.1 and C.3, the confounding bridge function is identified as the unique solution to the following equation:
\begin{equation}
\begin{split}
\nonumber
\frac{E[D_i\mid W_i, \mathbf{X}, \mathbf{A}]}
{E[1-D_i\mid W_i, \mathbf{X}, \mathbf{A}]}
=
E[q_2( Z_{j}, \mathbf{X}, \mathbf{A})\mid D_{i}=0, W_i, \mathbf{X},\mathbf{A}],
\end{split}
\end{equation}
and the AIT is identified by
\begin{equation}
\begin{split}
\nonumber
\tau_{ADT}(g)=\frac{1}{n}\sum_{i \in N_{n}}\left( 
E[\frac{D_i}
{E(D_i)}\sum_{j \in \mathcal{E} _{i}}\Delta Y_{j}]
-
E[\frac{1-D_i}
{E(D_i)}\sum_{j \in \mathcal{E} _{i}}q_2(Z_{j}, \mathbf{X}, \mathbf{A})\Delta Y_{j}] \right).
\end{split}
\end{equation}
\par
Let $h^*_2(W_{j}, \mathbf{X}, \mathbf{A})$ and $q^*_2(Z_{j}, \mathbf{X}, \mathbf{A})$ be arbitrary models for the true, unknown function $h_2(W_{j}, \mathbf{X}, \mathbf{A})$ and $q_2(Z_{j}, \mathbf{X}, \mathbf{A})$ for $i$ $\in$ $N_n$ and $j$ $\in$ $\mathcal{E} _{i}$. In order to describe our proposed doubly robust approach, consider the following two models:
\begin{enumerate}
  \item Model $\mathcal{M}_3$, in which $h^*_2(W_{j}, \mathbf{X}, \mathbf{A})$ = $h_2(W_{j}, \mathbf{X}, \mathbf{A})$ and Assumptions 2.2, C.1 and C.2 hold.
  \item Model $\mathcal{M}_4$, in which $q^*_2(Z_{j}, \mathbf{X}, \mathbf{A})$ = $q_2(Z_{j}, \mathbf{X}, \mathbf{A})$ and Assumptions 2.2, C.1 and C.3 hold.
  \end{enumerate}
The doubly robust DID estimand can be expressed as:
\begin{equation}
\begin{split}
\nonumber
\tau^{dr}_{AIT}=\frac{1}{n}\sum_{i \in N_{n}} 
E\left[\sum_{j \in \mathcal{E} _{i}}\left(\frac{D_i}
{E(D_i)}
-
\frac{q^*_2(Z_{j}, \mathbf{X}, \mathbf{A})(1-D_i)}
{E(D_i)}\right)
\left(\Delta Y_{j}-h^*_2(W_{j}, \mathbf{X}, \mathbf{A})\right)\right] .
\end{split}
\end{equation}
The following Proposition C.1 shows that our proposed doubly robust estimand recovers the ADT provided that at least one of models $\mathcal{M}_1$ and $\mathcal{M}_2$ is correctly specified.\par
$\mathbf{Proposition\ C.1}$
\\
If at least one of the models $\mathcal{M}_3$ and $\mathcal{M}_4$ is correctly specified, then $\tau^{dr}_{AIT}$ = $\tau_{AIT}$.

\end{document}